\documentclass[sigplan,10pt]{acmart}

\usepackage{tikz}
\usetikzlibrary{arrows.meta,positioning}
\usepackage{pgfplots}
\pgfplotsset{compat=1.18}
\usepackage{amsmath}
\usepackage{upgreek}
\usepackage[normalem]{ulem}
\usepackage{endnotes,microtype,xspace,fancyvrb,multirow}
\usepackage{xcolor}
\usepackage{xspace}
\usepackage{balance}
\usepackage{pifont}
\usepackage{anyfontsize}
\usepackage{booktabs}
\usepackage{caption}
\usepackage{listings}
\usepackage{soul}
\usepackage{textcomp}
\usepackage{enumitem}
\usepackage{array}
\usepackage{subcaption}
\usepackage{tabularx}
\usepackage{fontenc}
\usepackage{xparse}

\newcommand{\sys}{DeltaBox\xspace}
\newcommand{\agentfs}{DeltaFS\xspace}
\newcommand{\agentcr}{DeltaCR\xspace}
\newcommand{\agentstate}{DeltaState\xspace}
\newcommand{\dckpt}{deltaCheckpoint\xspace}
\newcommand{\drestore}{deltaRestore\xspace}

\newcommand{\statemgr}{StateManager\xspace}

\newcommand{\myparagraph}[1]{\par\addvspace{2pt plus 1pt minus .2pt}\noindent\textbf{#1}}

\newcommand{\fcdiffdm}{\texttt{FC\allowbreak-Diff\allowbreak+dm}\xspace}

\newcommand{\replaycp}{\texttt{replay\allowbreak+cp}\xspace}

\newcommand{\criucp}{\texttt{CRIU\allowbreak+cp}\xspace}

\definecolor{diaozi}{RGB}{93,49,49}
\definecolor{hightcode}{rgb}{1.0,0.13,0.32}

\copyrightyear{2026}
\acmYear{2026}
\setcopyright{cc}
\setcctype{by}
\acmConference[SIGOPS ATC 2026]{The 2026 ACM SIGOPS Annual Technical Conference}{November 15--18, 2026}{Hong Kong, China}
\acmBooktitle{The 2026 ACM SIGOPS Annual Technical Conference (SIGOPS ATC 2026), November 15--18, 2026, Hong Kong, China}
\acmDOI{10.1145/3828523.3857015}
\acmISBN{979-8-4007-2818-1/2026/11}

\acmBadgeR[https://www.acm.org/publications/policies/artifact-review-and-badging-current]{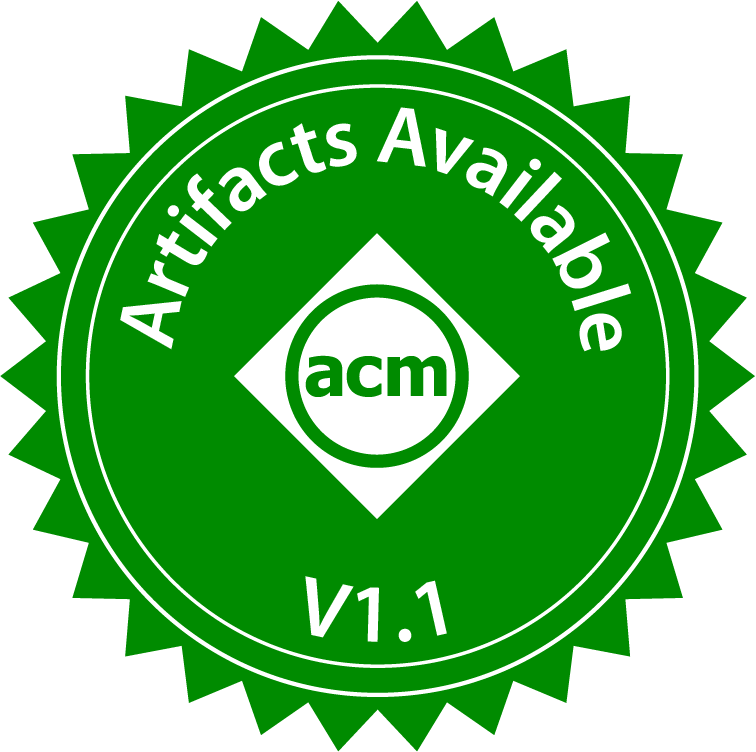}
\acmBadgeR[https://www.acm.org/publications/policies/artifact-review-and-badging-current]{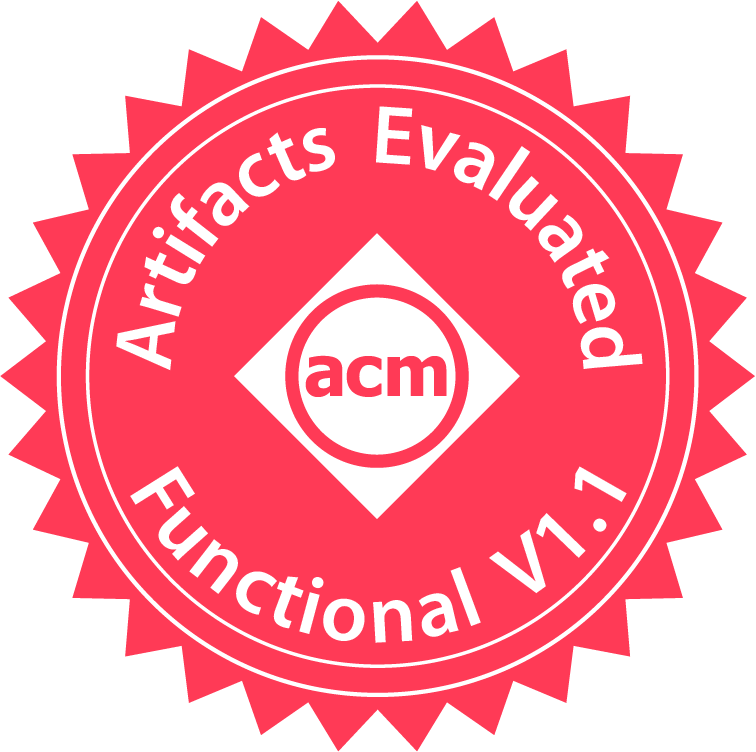}
\acmBadgeR[https://www.acm.org/publications/policies/artifact-review-and-badging-current]{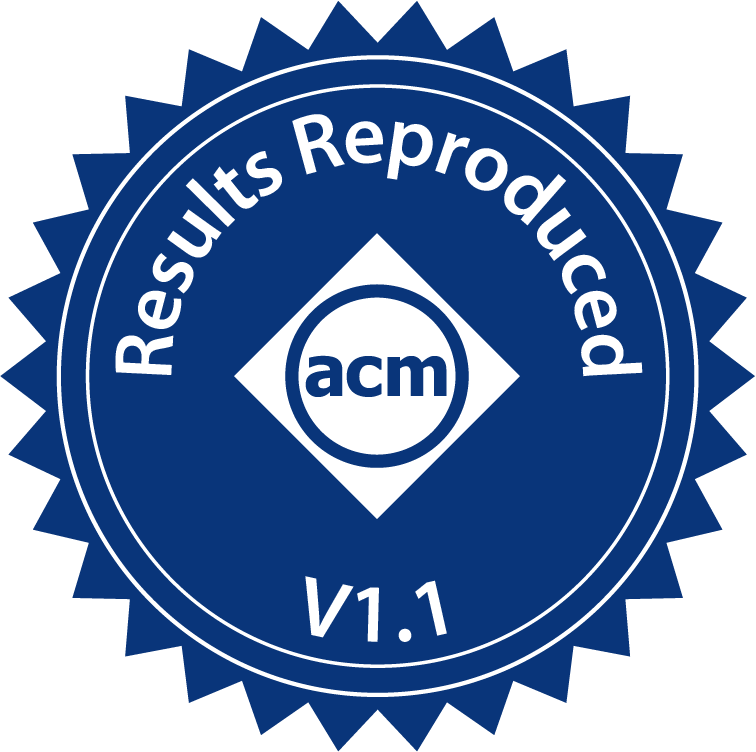}

\begin{document}

\title{DeltaBox: Millisecond-Scale Sandbox Checkpoint and Rollback for Stateful AI Agents}

\author{Yunpeng Dong}
\orcid{0009-0005-5434-0435}
\affiliation{\institution{Shanghai Jiao Tong University}
\city{shanghai}
\country{China}}
\email{dongyunpeng@sjtu.edu.cn}

\author{Jingkai He}
\orcid{0009-0005-9024-7588}
\affiliation{\institution{Shanghai Jiao Tong University}
\city{shanghai}
\country{China}}
\email{hjk020101@sjtu.edu.cn}

\author{Shiqi Liu}
\orcid{0009-0008-8388-6711}
\affiliation{\institution{Shanghai Jiao Tong University}
\city{shanghai}
\country{China}}
\email{sjtu\_666@sjtu.edu.cn}

\author{Yuze Hou}
\orcid{0009-0009-6644-4601}
\affiliation{\institution{Shanghai Jiao Tong University}
\city{shanghai}
\country{China}}
\email{hyzhyzsh@sjtu.edu.cn}

\author{Dong Du}
\orcid{0000-0002-7945-8430}
\affiliation{\institution{Shanghai Jiao Tong University}
\city{shanghai}
\country{China}}
\email{dd\_nirvana@sjtu.edu.cn}

\author{Zhonghu Xu}
\orcid{0000-0001-7872-2992}
\affiliation{\institution{Huawei Technologies Co., Ltd.}
\city{shenzhen}
\country{China}}
\email{xuzhonghu1@huawei.com}

\author{Si Yu}
\orcid{0009-0002-5907-4306}
\affiliation{\institution{Huawei Technologies Co., Ltd.}
\city{shenzhen}
\country{China}}
\email{si.yu@huawei.com}

\author{Baochuan Yang}
\orcid{0009-0008-5407-1273}
\affiliation{\institution{Huawei Technologies Co., Ltd.}
\city{shenzhen}
\country{China}}
\email{bc.y@huawei.com}

\author{Yubin Xia}
\orcid{0000-0001-6558-5298}
\affiliation{\institution{Shanghai Jiao Tong University}
\city{shanghai}
\country{China}}
\email{xiayubin@sjtu.edu.cn}

\author{Haibo Chen}
\orcid{0000-0002-9720-0361}
\affiliation{\institution{Shanghai Jiao Tong University}
\city{shanghai}
\country{China}}
\email{haibochen@sjtu.edu.cn}

\begin{abstract}
Emerging LLM-powered AI agent workloads, such as test-time tree search and reinforcement learning (RL), require high-frequency state exploration, relying on rapid checkpoint/restore (C/R) of the complete sandbox state, including files and process state.
Existing mechanisms duplicate the entire state, causing hundreds of milliseconds to seconds of latency per C/R, which severely bottlenecks deep search and large-scale fan-outs. 

We observe that consecutive sandbox states between checkpoints differ only marginally.
Therefore, instead of full duplication, a sandbox should \emph{only duplicate the changes between consecutive checkpoints} (\textbf{Key Insight}), yet the lack of OS support makes this hard to realize.
We propose a new OS-level abstraction, \emph{\agentstate}, that delivers change-based C/R for agents through co-designed OS mechanisms.
First, \emph{\agentfs} checkpoints file state by freezing the writable layer and inserting a new one, turning updates into copy-on-write and rollback into a layer switch.
Second, \emph{\agentcr} quickly restores process state by forking from a frozen template process. A coordinator pairs the two so each checkpoint is one consistent (filesystem, memory) state.
Building on these mechanisms, we present \sys, an agent sandbox with millisecond-level C/R.
On SWE-bench, \sys hides its checkpoint work (10.83\,ms) under inference and rolls back in a mean 1.86\,ms, with the same warm-template fork driving low-cost RL rollout fan-out.
\sys is open-sourced at \url{https://github.com/delta-box}.

\end{abstract}

\renewcommand{\shortauthors}{Dong et al.}


\maketitle

\section{Introduction}
\label{sec:intro}

LLM-powered agents now automate complex, multi-step tasks, from code repair~\cite{swebench} to web and app navigation~\cite{webarena,zhang2025mobiagentsystematicframeworkcustomizable}.
An agent runs an iterative loop: it generates an action~\cite{yao2023react}, executes it in a sandbox~\cite{e2b}, observes the outcome, and refines its next step.
To push capability further, agents increasingly scale test-time compute~\cite{snell2024scaling} through tree search such as Monte Carlo Tree Search (MCTS)~\cite{lats} and Tree of Thoughts~\cite{NEURIPS2023_271db992}, exploring multiple candidate paths, checking them against execution feedback, and backtracking from failures.

In early text-based tasks (e.g., HotPotQA), backtracking cost almost nothing: the agent simply truncated its prompt history.
A coding agent, by contrast, acts on real OS state~\cite{openclaw} through commands like \texttt{rm}, \texttt{pip install}, and \texttt{sed}.
These side effects persist on disk and cannot be undone by editing history. Reverting them requires checkpointing and restoring the complete sandbox state.
Because tree search backtracks constantly, this expensive rollback repeatedly hits the critical path---an infrastructure bottleneck.

The sandbox state encompasses two tightly coupled dimensions: durable filesystem state and ephemeral process state.
These dimensions must be captured and restored jointly to prevent state divergence or context loss.

Test-time search stresses checkpoint/restore (C/R) along \emph{two axes} at once: horizontally, Best-of-N sampling spreads the inference budget across many parallel trajectories, each needing a fast initial clone~\cite{bon-scaling,snell2024scaling}; vertically, tree search such as MCTS backtracks to arbitrary historical nodes, demanding fine-grained C/R of intermediate states.

Beyond test-time inference, RL training of agent policies is a second pressure point~\cite{10.1145/3779212.3790172,shao2024deepseekmathpushinglimitsmathematical,NEURIPS2025_a4277440}.
Each training step must spin up $N$ \emph{independent} sandboxes from the same warm starting state, run them as rollouts, and tear them down.
Today's systems rebuild this warm state per rollout with a fresh Docker container or a microVM snapshot-restore, both in the hundreds-of-milliseconds to seconds regime. With $N$ in the tens to low hundreds, fork latency bounds training throughput.

Existing systems manage filesystem and process state separately, achieving C/R by duplicating the entire state into a checkpoint image~\cite{daytona,ftsandbox,criu,e2b-checkpoint,firecracker}.
Full duplication is acceptable for serverless cold starts~\cite{faasnap, catalyzer, reap, 10.5555/3767901.3767929}, but prohibitively slow for agents that checkpoint and roll back frequently on the critical path.
This creates three obstacles for agent workloads:

\myparagraph{Challenge-1: Existing durable filesystem C/R is \emph{coarse-grained and disruptive}.}
A significant challenge is how to efficiently checkpoint and restore the durable filesystem state of an agent sandbox, including repository edits, installed packages, caches, and temporary files.
Naive file-level copying (e.g., \texttt{copytree}) scales with the directory size, Docker commits scale with the mutated layer, and VM/block snapshots~\cite{firecracker} operate at a coarse whole-VM granularity.
These approaches are tolerable for occasional resets, but too slow for per-step rollback in MCTS or large fan-out in RL.
Overlayfs appears to be a natural alternative because each checkpoint could demote the current writable upper into a read-only lower and install a fresh upper, turning later writes into copy-on-write (CoW).
However, stock Linux overlayfs freezes its layer stack at mount time: changing the stack requires \texttt{umount}/\texttt{mount}, which is impossible while an agent keeps files or working directories open and also flushes dentry/inode caches~\cite{overlayfs}.
Moreover, ordinary overlayfs copy-up can duplicate an entire inherited file for a small edit, causing file-size-proportional write amplification.

\myparagraph{Challenge-2: \emph{Whole-state dump and materialization} limit process C/R efficiency.}
An agent's process state is not just heap bytes: it includes the Python interpreter, loaded indexes, tool context, open descriptors, and SDK helper threads that make a sandbox warm.
Existing process C/R exposes a whole-state dump/materialize interface: checkpoint serializes the warm footprint, and restore reconstructs it, even when the agent changed little since the previous step.
This model is acceptable for serverless cold-start systems that optimize restore and treat snapshot creation as offline work~\cite{10.1145/3694715.3695966, 10.5555/3767901.3767929, catalyzer, 10.1145/3503222.3507732, 280716, 10.1145/3617232.3624871, trenv-x, 10.1145/3694715.3695947}, but mismatched with agents, which checkpoint after nearly every action and restore on every backtrack or fork-out.
Whole-VM snapshots pause and copy unrelated guest state (E2B reports ${\sim}4$\,s per GiB~\cite{e2b-checkpoint}), while process-level CRIU~\cite{criu} avoids VM scope but still serializes and reloads process pages, file mappings, and descriptors.

\myparagraph{Challenge-3: \emph{Coupled sandbox state} requires explicit consistency management.}
Even if filesystem and process C/R are optimized in isolation, agents cannot restore the two dimensions independently.
Process memory contains open descriptors, mmap'd files, Python import caches, tool outputs, and assumptions about packages or tests on disk; the filesystem records the edits and side effects that this memory will continue to use after rollback.
A naive composition of overlayfs and CRIU can capture the two dimensions at different logical times, register a partially completed checkpoint after a failure, or resume a process before its target filesystem is installed.
Agent sandboxes therefore need explicit consistency management for an atomic, failure-aware (filesystem, process) state pair, not two unrelated snapshots.

This paper observes that consecutive checkpoints in agent workloads differ only marginally, with only a few new files or modified memory pages between steps.
Therefore, instead of duplicating the entire state, a sandbox should \emph{only duplicate the changes between consecutive checkpoints} (\textbf{Key Insight}).

To this end, we present \sys, an OS-level rollbackable sandbox tailored for stateful agents.
\sys achieves millisecond-level checkpoint/rollback
through a new OS abstraction, \agentstate,
which treats the filesystem and process memory as a transactional, change-based state pair.
We introduce two co-designed OS mechanisms in \sys, coordinated by a \statemgr:

\begin{figure}[!t]
  \centering
  \includegraphics[width=\columnwidth]{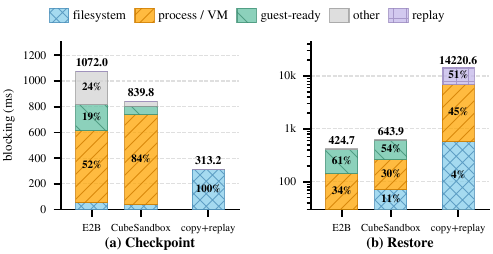}
  \caption{\textbf{Per-event checkpoint/restore cost of existing approaches on SWE-Search}, decomposed into filesystem, process/VM, and other phases (restore on a log axis).}
  \label{fig:cr-cost}
\vspace{-16pt}
\end{figure}

Firstly, to support efficient filesystem C/R, we propose \textbf{\agentfs}. 
\agentfs extends overlayfs with \emph{runtime layer switching}, which reconfigures the layer stack without unmounting or closing open file descriptors. 
This makes a checkpoint a metadata-only layer insertion that preserves history, while restore selects a recorded layer-stack configuration in $O(1)$ time. Reference-counted generations keep cached dentries and inodes safe while old stacks drain.
To keep the switch transparent to a live agent, \agentfs adds a \emph{lazy switch for open files}: per-inode version tags detect open files or mmap'd regions resolved before the checkpoint and re-resolve/copy them up on the next write. 
Finally, \agentfs uses \emph{XFS reflink} under overlayfs copy-up, so unchanged extents remain shared across renamed generations and file updates pay only block-level \emph{CoW} rather than file-size-proportional duplication.

Secondly, to C/R process memory state efficiently, we design \textbf{\agentcr}. 
\agentcr builds process checkpoints around a \emph{warm-template} fast path: between actions, the agent inline-forks a frozen template while \agentcr asynchronously dumps that template with CRIU, giving each checkpoint both a low-millisecond restore target and a durable fallback, with the work hidden inside the LLM I/O window. 
On a template hit, \emph{fork-based restore} spawns the restored agent by copying page tables rather than memory pages; on a miss, \emph{CRIU lazy-pages restore} reconstructs the process on demand and reinserts it into the pool. 
The template serves the common single-threaded agent at fork speed, while the CRIU image captures the whole process tree, extending the same protocol to multithreaded and multi-process agents.
To keep deep search bounded, \agentcr \emph{skips read-only nodes} to avoid snapshots for cheap-to-recompute actions, and \emph{reachability-aware garbage collection} reclaims only snapshots the search can no longer select. 
Finally, we propose \emph{async-warm} to absorb post-fork CoW faults off the critical path. A \emph{network proxy daemon} keeps LLM sockets and SDK threads outside the template so fork is safe. For RL fan-out, \sys implements batched \texttt{fork\_n} on top of the same template protocol.

Thirdly, to keep memory and file views consistent, \sys adds a \textbf{\statemgr} and a coupled C/R protocol.
The \statemgr maintains a snapshot index tree mirroring the search tree, where each node records the target overlay layer-stack configuration, CRIU dump path, and live template when available.
On checkpoint, the boundary between actions lets \agentcr capture the post-action process and \agentfs freeze the matching filesystem generation before the node becomes visible; on restore, \sys switches the filesystem first and resumes the restored process only afterward.
This ordering keeps memory and file views mutually consistent, and serves rollback, evaluation-time test isolation, and batched RL fork-out.

We implement \agentfs as an overlayfs kernel module and \agentcr as a userspace framework, integrating both in \sys, an agent sandbox based on a Firecracker microVM.
End-to-end evaluations show that a checkpoint's 10.83\,ms of work is hidden under inference (zero agent-perceived blocking), while template-fork restore completes in a mean 1.86\,ms. The CRIU lazy-pages fallback after template eviction restores in 9.29\,ms.
On SWE-bench MCTS workloads, \sys reduces state-management overhead from 39--51\% of total time on the baselines to 1--2\%.
For RL training fan-out, \sys's batched \texttt{fork\_n} forks up to 64 rollouts from one warm template at 0.55--5.1\,ms per child.
End-to-end fan-out is $10.4\times$--$24.7\times$ faster than CubeSandbox and E2B.

\section{Background and Motivation}
\label{sec:background}

\begin{table*}[t]
\centering
\caption{\textbf{Comparison of sandbox state management approaches for agents.}}
\vspace{-5pt}
\label{tab:comparison}
\fontsize{8pt}{9pt}\selectfont
\begin{tabular}{@{}lcccccccc@{}}
\toprule
\textbf{Approach} & \textbf{Ckpt} & \textbf{Restore} & \textbf{Write} & \textbf{Mem.} & \textbf{FS} & \textbf{Process} & \textbf{Arbitrary} & \textbf{Low control} \\
 & \textbf{Latency} & \textbf{Latency} & \textbf{Amplif.} & \textbf{Sharing} & \textbf{State} & \textbf{State} & \textbf{Rollback} & \textbf{overhead} \\
\midrule
Git worktree & 100\,ms--1\,s & 100\,ms--1\,s & $O(\text{files})$ & \ding{55} & \ding{51} & \ding{55} & \ding{51} & \ding{51} \\
copytree & 100\,ms--10\,s & 100\,ms--10\,s & $O(\text{dir})$ & \ding{55} & \ding{51} & \ding{55} & \ding{51} & \ding{51} \\
Docker commit + restart & 50\,ms--10\,s & 1--10\,s & $O(\text{layer})$ & \ding{55} & \ding{51} & \ding{55} & \ding{51} & \ding{51} \\
Btrfs/LVM snapshot & 10--100\,ms & 10\,ms--1\,s & Low & \ding{55} & \ding{51} & \ding{55} & \ding{51} & \ding{51} \\
Firecracker VM snapshot & 200\,ms--2\,s & 120--700\,ms & $O(\text{VM})$ & \ding{55} & \ding{55} & \ding{51} & \ding{51} & \ding{51} \\
\midrule
DSec~\cite{deepseek-v4} & --- & replay & $O(\text{log})$ & \ding{55} & \ding{55} & \ding{55} & \ding{55} & \ding{51} \\
CubeSandbox~\cite{cubesandbox} & 839.8\,ms\textsuperscript{\S} & 643.9\,ms\textsuperscript{\S} & $O(\text{VM})$ & \ding{51} & \ding{51} & \ding{51} & \ding{51} & \ding{55} \\
E2B~\cite{e2b-checkpoint} & 1072.0\,ms\textsuperscript{\S} & 424.7\,ms\textsuperscript{\S} & $O(\text{VM})$ & \ding{55} & \ding{51} & \ding{51} & \ding{51} & \ding{55} \\
\midrule
\textbf{\sys (ours)} & \textbf{10.83\,ms} & \textbf{1.86\,ms\textsuperscript{\P}} & \textbf{$O(\text{4KB})$} & \textbf{\ding{51}} & \textbf{\ding{51}} & \textbf{\ding{51}} & \textbf{\ding{51}} & \textbf{\ding{51}} \\
\bottomrule
\end{tabular}

\vspace{2pt}
{\footnotesize \textsuperscript{\S} CubeSandbox v0.3.0 (CubeCoW) and self-hosted E2B, measured locally per event on our SWE-bench MCTS workload (\autoref{sec:eval}).\quad
\textsuperscript{\P} Fast path (fork): 1.86\,ms. Slow path (CRIU lazy-pages fallback): 9.29\,ms. See \autoref{tab:latency} for component breakdowns.}
\vspace{-12pt}
\end{table*}

\subsection{Search and Exploration for AI Agents}
\label{sec:bg-search}

Modern LLM-based agents increasingly rely on search and exploration, whether tree-structured or parallel sampling, to solve complex tasks such as math and code~\cite{swebench}.

\myparagraph{MCTS.}
MCTS is a tree-search algorithm that cycles through selection, expansion, evaluation, and backpropagation~\cite{kocsis2006uct, browne2012mctssurvey}. LATS~\cite{lats} adapts it to LLM agents, using UCT-guided selection to pick the most promising node and then generating and executing candidate actions in a sandbox.
SWE-Search~\cite{swe-search} specializes this loop to SWE-bench code repair, improving the success rate over single linear generation.

\myparagraph{Best-of-N.}
BoN runs $N$ independent trajectories from the same initial state and keeps the best by execution feedback~\cite{bon-scaling,snell2024scaling}, i.e., $N$ clones of the initial environment.

\myparagraph{RL training.}
RL post-training fans out rollouts (i.e., inference and agent exploration) at every training step. Chain-based algorithms such as GRPO~\cite{shao2024deepseekmathpushinglimitsmathematical} launch $N$ independent rollouts from the same initial state and score the completed trajectories. Tree-based variants such as Tree-GRPO~\cite{ji2026tree} instead branch additional rollouts from intermediate states of existing trajectories, sharing the common prefix instead of re-executing it. Either way the fan-out recurs at every training step, so sandbox cloning latency directly gates training throughput. Tree-based RL additionally requires forking from arbitrary mid-trajectory states.

\subsection{Agent Sandbox and Coupled Checkpoint/Restore}
\label{sec:bg-bottleneck}

Sandbox runtimes~\cite{firecracker, catalyzer} typically provide two critical primitives for agent exploration---\emph{checkpoint} and \emph{rollback}---letting agents execute code in isolation, capture milestone snapshots, and safely revert after failures.
RL fan-out (\autoref{sec:bg-search}) uses the same mechanism: the runtime checkpoints one warm environment and restores it $N$ times, once per rollout.

While most modern agent frameworks utilize some form of checkpoint and rollback, their implementations are often inefficient or incomplete.
For example, SWE-agent~\cite{sweagent} and OpenHands~\cite{openhands} run the agent control logic on the host and reset the in-sandbox repository using Git.
However, this approach fails to checkpoint the running process,
leading to a loss of the execution context (e.g., debugging history).
To preserve this context, current frameworks interact with the environment via a persistent daemon process inside the sandbox (e.g., E2B's envd~\cite{e2b}).
In this setup, a rollback only undoes file edits, leaving the process state desynchronized.
Recovering the true process state requires \emph{replaying every prior action} from the initial state, which is a latency-heavy operation proportional to the search depth and may severely bottleneck the critical path of backtracking search.

\myparagraph{Coupled C/R matters.}
The state of a sandbox spans two interdependent dimensions:
the \emph{filesystem} (FS) (working directories, installed packages)
and the \emph{process memory} (execution context, open file descriptors).
A partial rollback inevitably desynchronizes them.
If only the FS is rolled back, the still-live process retains memory of file contents that no longer exist on disk.
Conversely, if only the process is rolled back, its restored memory operates on files that have been mutated beyond its expected state.
Either mismatch breaks the strict consistency required for correct tree search (\autoref{sec:eval-correctness}), necessitating a truly coupled C/R mechanism.

\subsection{Why Existing Approaches Fall Short}
\label{sec:bg-existing}

We summarize existing agent sandboxes in \autoref{tab:comparison}.

\myparagraph{Checkpoint latency matters in agents.}
Although sandboxing is well studied in contexts like serverless computing, existing optimizations predominantly focus on minimizing \emph{restore} latency to mitigate cold starts~\cite{catalyzer, firecracker, trenv-x, sock}.
However, in agents, the \emph{checkpoint} operation also sits squarely on the critical path: \emph{an agent cannot proceed until its state is safely snapshotted}.
This creates a new performance requirement that prior systems overlook. 

\myparagraph{Efficient coupled C/R is difficult.}
Furthermore, existing systems manage filesystem and process states in isolation, failing to provide the coupled C/R efficiency required by agents, as shown in \autoref{tab:comparison}.
\emph{Filesystem-only} methods (e.g., Git worktree~\cite{claudecode-worktrees}) version files but discard process memory, forcing expensive agent restarts.
Conversely, \emph{Process/VM-only} methods (e.g., Firecracker snapshots~\cite{firecracker}, CRIU) capture memory but fail to coordinate with the underlying block device, necessitating separate, uncoupled filesystem snapshots.

\begin{figure}[t]
  \centering
  \includegraphics[width=0.99\columnwidth]{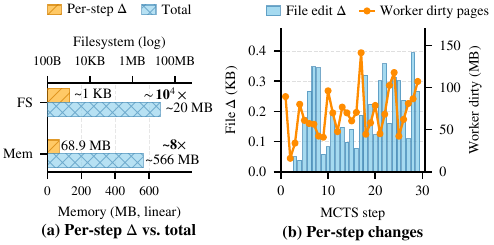}
  \caption{\textbf{Each MCTS step changes only a tiny fraction of the total sandbox state.} \textbf{(a)}~Mean per-step change vs.\ total state for the filesystem and process memory. \textbf{(b)}~Per-step filesystem edit size and worker dirty pages.
}
  \label{fig:motiv}
\vspace{-12pt}
\end{figure}

\myparagraph{A quantitative analysis.}
\autoref{fig:cr-cost} breaks down the per-event C/R latency we measured for self-hosted E2B~\cite{e2b-checkpoint} (incremental snapshots), CubeSandbox~\cite{cubesandbox}, and a \texttt{copytree+replay} baseline.
Both VM-based sandboxes spend little time on the filesystem (at most 11\% of any operation), with most time spent on VM state management and guest readiness.
CubeSandbox spends 84\% of its 839.8\,ms checkpoint creating the VM memory snapshot, and E2B's 1072.0\,ms checkpoint pauses the VM, snapshots it, and resumes it before the agent can respond again.
On restore, waiting for the agent's first response after the VM resumes takes 61\% of E2B's 424.7\,ms (85\% including envd re-initialization) and 54\% of CubeSandbox's 643.9\,ms.
The \texttt{copytree+replay} baseline keeps no process checkpoint: copying files takes only 4\% of its 14.2\,s restore, while restarting the agent and reloading its indexes (45\%) and re-executing prior actions (51\%) take the rest.
These costs make repeated state transitions expensive, even though each MCTS step changes only a small fraction of sandbox state (\autoref{fig:motiv}).

\subsection{Insight: Delta-Based Checkpoint and Restore}

We analyzed state changes in SWE-Search MCTS trajectories (\autoref{fig:motiv}).
Per step, process memory and filesystem deltas average only 12.2\% and ${\sim}0.005\%$ of their total state, respectively (\autoref{fig:motiv}(a)), consistently across rounds and workloads.
Small state deltas persist across MCTS steps (\autoref{fig:motiv}(b)).

This observation leads to our core idea:
rather than duplicating the entire sandbox state, an efficient agent sandbox should \emph{only capture and restore the state deltas between consecutive checkpoints} (\textbf{Key Insight}), bypassing the I/O and memory-copy overheads of full-system C/R.

\begin{figure*}[t]
\centering
\includegraphics[width=\linewidth]{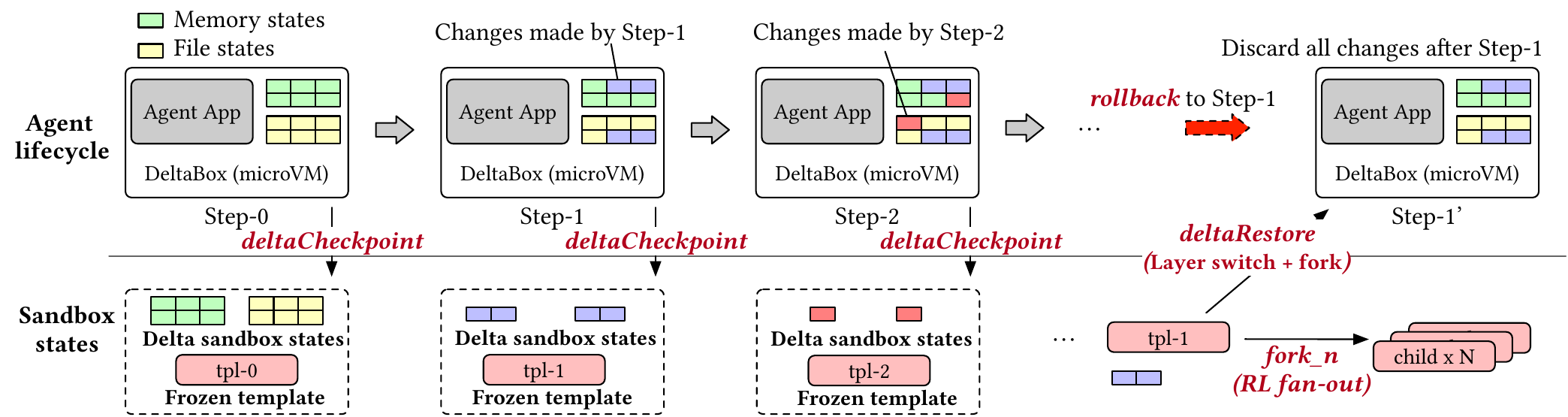}
\caption{\textbf{Design Overview}. The agent worker runs inside the sandbox, so every step is C/R-protected. From one diff-based checkpoint/restore primitive, \sys serves two workloads: millisecond \emph{rollback} to a prior node for tree search, and \emph{$N$-way fork-out} from a frozen template for parallel fan-out.}
\label{fig:design-overview}
\vspace{-16pt}
\end{figure*}

\section{Overview}
\label{sec:overview}

Based on the insight, we present \sys, a new agent sandbox realizing millisecond-scale C/R.

\subsection{System Architecture}

\sys extends existing sandboxes (e.g., microVMs) with two key components: \agentfs and \agentcr.

\myparagraph{{\agentfs}: Delta-based filesystem.}
A fundamental challenge for \sys is efficient C/R of filesystem state.
A natural approach is to apply CoW to the filesystem: keep existing files read-only and copy them only when modified, so each checkpoint need only record changes since the previous checkpoint.

Overlayfs~\cite{overlayfs} offers the closest realization of this idea.
It provisions a writable upper layer for apps while maintaining a read-only lower layer that encapsulates the base sandbox image. 
However, overlayfs suffers from a critical limitation:
its layer stack is \emph{fixed and static} at mount time.

To resolve this, we introduce \agentfs, a new filesystem extending overlayfs that supports dynamic, runtime overlay reconfiguration.
When an agent completes a task step and requests a checkpoint, \agentfs does not duplicate the underlying files.
Instead, it dynamically freezes the current writable layer into a read-only state and seamlessly inserts a new writable upper layer on top of it.
Consequently, all subsequent file updates trigger CoW to this new upper layer.
Through this mechanism, a checkpoint operation is reduced to a lightweight layer-insertion, and a rollback becomes a simple layer-removal, aligning with our diff-based C/R insight.
\agentfs further uses reflink-enabled XFS, whose block-level CoW keeps copy-up cost proportional to the blocks actually written (\autoref{sec:design-hotswitch}).

\myparagraph{{\agentcr}: Delta-based process checkpoint/restore.}
Process-level state management has seen extensive prior research.
Methods like CRIU~\cite{criu} and \texttt{fork()}~\cite{catalyzer, 288627} are widely used.
However, applying them directly to agent workloads presents severe challenges.
While recent efforts have optimized CRIU's restore latency, its checkpoint latency, even using incremental dumps~\cite{criu-incremental-dump},
remains 100--1,000\,ms.
While \texttt{fork()} offers millisecond-scale cloning,
it introduces its own limitations:
(1) it requires the parent process (template) to remain resident in memory, scaling poorly for agent exploration;
(2) CoW page faults add latency right after each restore;
and (3) both CRIU and \texttt{fork()} are only safe if the process holds no long-lived external state.
Standard Python LLM SDKs (e.g., OpenAI) violate this by holding long-lived connections.

To this end,
\sys introduces \agentcr, a hybrid process C/R mechanism that synthesizes the durability of CRIU with the speed of \texttt{fork()}.
During a checkpoint, \agentcr simultaneously executes an asynchronous CRIU incremental dump and a template-creating \texttt{fork()}, 
hiding both operations within the LLM I/O window.
During a restore, \sys prioritizes the fast-path by forking the resident template.
To bound memory costs, \agentcr employs a template pool, transparently falling back to CRIU's lazy-pages restore if the template has been evicted.
\agentcr includes an optional async-warm thread that absorbs post-fork CoW faults off the critical path (\autoref{sec:design-mem}), alongside a network proxy daemon that keeps persistent I/O connections outside the agent process so the frozen template remains forkable.

\myparagraph{Coordination and optimizations.}
To ensure atomic, coupled state transitions,
\sys employs a two-tiered \statemgr to orchestrate \agentfs and \agentcr, maintaining each checkpoint as one \agentstate: a consistent, transactional (filesystem, memory) pair.
The global Sandbox Controller runs on the host, maintaining a snapshot index tree isomorphic to the search tree,
where each node records its CRIU dump path and overlay layer-stack configuration.
Simultaneously, the Guest Daemon executes the local C/R sequences inside the VM,
keeping state management off the host's critical path (\autoref{sec:overview-deploy}). 

\sys also introduces agent-specific optimizations to further minimize overhead.
First, a \emph{semantic-aware lightweight skip} mechanism avoids unnecessary snapshots.
Second, \sys utilizes a \emph{reachability-aware garbage collector},
to reclaim unused checkpoints.

\subsection{System Workflow}

\autoref{fig:design-overview} shows the overall workflow of \sys.
\sys efficiently supports two critical workloads: millisecond-scale rollback to a prior node for tree search, and rapid $N$-way fork-out from a frozen template for parallel RL training.

\subsubsection{Coupled \dckpt and \drestore}
\label{sec:overview-checkpoint}

We introduce the workflow of two primitives.

\myparagraph{\dckpt.}
A checkpoint operation is orchestrated by the \statemgr and executed entirely off the host's critical path by the Guest Daemon inside the sandbox.
First, the daemon captures the process state via \agentcr: the agent executes a \texttt{fork()} at a quiescent point to create a frozen template, while a CRIU incremental dump is asynchronously submitted.
Concurrently, the daemon triggers the \agentfs \texttt{ioctl} to freeze the current FS upper layer and install a fresh writable layer.
The agent remains perfectly quiescent across both operations.
This guarantees that every \dckpt registers a strictly consistent (FS, memory) state delta.
Because this entire sequence executes concurrently while the agent awaits its LLM response, the checkpoint cost is hidden behind inference latency.

\myparagraph{\drestore.}
A restore operation seamlessly reverses this process.
On rollback, the Sandbox Controller instructs the Guest Daemon to discard the current active agent.
The daemon then uses the \agentfs \texttt{ioctl} to dynamically revert the FS layer stack to the target checkpoint's exact configuration.
Simultaneously, \agentcr rebuilds the process memory.
If the target's frozen template is still resident in the pool, \agentcr executes a millisecond \texttt{fork()} (the fast path); otherwise, it falls back to a CRIU lazy-pages restore.
The new agent process then resumes execution from the exact instruction immediately following the original checkpoint, entirely unaware of the rollback, with its FS and memory perfectly synchronized.

\subsubsection{Fork-out}
\label{sec:overview-forkout}

Fork-out composes the same two primitives differently: rather than restoring one past state, it instantiates $N$ children from one checkpoint, chain-based RL from the warm initial state and tree-based RL from any mid-trajectory checkpoint (\autoref{sec:bg-search}).
The \statemgr takes a \dckpt at the source state, then forks the checkpoint's frozen template $N$ times in one batched \texttt{fork\_n} request, each child receiving a private filesystem view over the frozen layers (\autoref{sec:design-forkout}).
Each child starts with its memory CoW-shared with its siblings, at the same millisecond-scale per-child cost as a fast-path \drestore (\autoref{sec:eval-throughput-rl}).

\subsection{Branch-Level Containment Model}
\label{sec:containment}
A \emph{branch} is one search path or forked rollout of a task, and \sys isolates branches within a single VM at process and file granularity: each branch runs in a separate process with a private writable overlay upper, while frozen ancestor layers and read-only base layers are shared.
The guest kernel, network proxy, and files outside the \agentfs-managed task repository are shared and not checkpointed.
Rolling back one branch thus leaves its siblings untouched.
Existing agent sandboxes (e.g., E2B~\cite{e2b} and CubeSandbox~\cite{cubesandbox}) usually snapshot and clone whole VMs, so each branch typically runs in its own VM, whereas \sys isolates many branches inside one.

\myparagraph{External effects.}
\sys rolls back only state inside the sandbox: the network proxy keeps communication continuous across checkpoints and restores (\autoref{sec:design-proxy}) but undoes nothing remotely.
Stateless LLM requests and idempotent operations (e.g., package downloads)~\cite{288554} are simply reissued after a restore, whereas stateful remote operations (e.g., database writes) need cooperation beyond the sandbox~\cite{10.1145/568522.568525}, such as service-side transactions, deferred outputs, or compensating actions.

\section{Detailed Design}
\label{sec:design}

\subsection{\agentfs: Dynamic Overlay Filesystem}
\label{sec:design-hotswitch}

\begin{figure}[t]
\centering
\includegraphics[width=\linewidth]{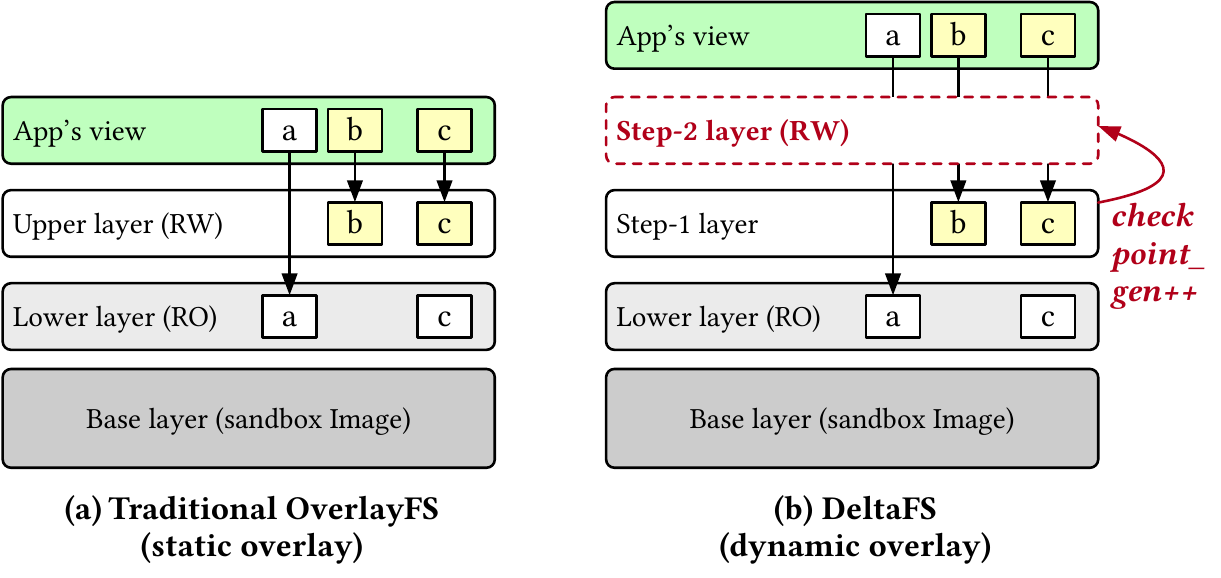}
\caption{\textbf{\agentfs architecture.} (a) Standard overlayfs has a static layer stack. (b) \agentfs reconfigures it at runtime: a checkpoint inserts a fresh writable upper, and a rollback switches to the target configuration; \texttt{checkpoint\_gen} stamps each switch with a new generation.}
\label{fig:design-fs}
\vspace{-16pt}
\end{figure}

Overlayfs fixes its layer stack at mount time. With live file references, remounting either fails or leaves the agent on a stale stack, and releasing those references means restarting the agent worker (0.4--17\,s in our workloads).
As shown in \autoref{fig:design-fs}, \agentfs lifts
this restriction and supports dynamic layers, switching the stack in 0.19\,ms (\autoref{tab:latency}) while the agent keeps its process and descriptor state.

\myparagraph{Interface.}
To expose this dynamic capability, \agentfs introduces a \texttt{hotswitch} ioctl that reconfigures the overlay layer stack in place to a target state: a writable upper layer above an ordered chain of read-only lower layers.

During a checkpoint, if the current upper is dirty, the \statemgr uses this interface to sink it into a frozen lower layer and install a new writable upper layer above it.
Consequently, files in the prior layer remain reachable, but any subsequent modifications trigger CoW to the new upper layer,
isolating the state delta.
The restore path is even simpler: the \statemgr invokes the ioctl to specify the exact lower layer chain of the target checkpoint and installs a fresh writable upper layer above it.
Because this process is purely a layer reconfiguration, it takes $O(1)$ time regardless of how far back in the search tree the target checkpoint lies.
Path lookups still walk the active stack, so their latency grows with its depth (\autoref{fig:war}(c)). Since only dirty checkpoints add a layer, the stack stays within the search-tree depth (${\le}15$ in our workloads).
The ioctl is entirely synchronous; once it returns, subsequent path resolution observes the new stack; already-open descriptors are handled by the lazy rebinding mechanism below.

\myparagraph{Layer lifetime.}
A hot-switch completes without freezing the filesystem, so open files and in-flight lookups may still reference the old layer stack after the ioctl returns.
\agentfs therefore reference-counts each retired stack: a cached overlay entry holds a reference to the stack it was resolved against, and the kernel frees a retired stack only after its last reference drops and an RCU grace period has elapsed.

\myparagraph{Open files across checkpoint and restore.}
\label{sec:design-lazy}
Files opened prior to a checkpoint present a challenge: their FDs may point to backing files within an outdated layer stack.
\agentfs resolves this via a generation-based lazy rebinding mechanism.
Each inode caches the \texttt{checkpoint\_gen} at which it was last resolved.
During any file access, the kernel compares this cached generation against the current global filesystem generation.
If the generations match, the access proceeds directly via the fast path with near-zero overhead.
If they differ, the kernel re-resolves the inode against the new layer stack: read-side accesses only refresh stale references and keep reading from read-only layers, while write-side accesses first detach the descriptor from its stale backing and redirect it to the current writable upper, copying file data when needed.
Both paths update the inode's generation and backing under a per-inode mutex.
Furthermore, if concurrent threads race to copy up the same file, the loser simply reuses the already-created upper layer rather than retaining its stale backing file.

\myparagraph{Descriptor semantics.}
A rollback replaces the filesystem view underneath live file descriptors.
\agentfs makes this deterministic. After a
checkpoint, writing through a stale descriptor copies up from its frozen
backing into the current upper, so checkpointed history remains immutable.
After a restore, a descriptor whose path exists in the target state rebinds
reads and later copy-ups to that state.
If the file existed only in a discarded
upper,
the descriptor continues through an anonymous backing copy and never
resurrects the old path.

\myparagraph{XFS reflink.}
\label{sec:design-reflink}
Layer switching and lazy rebinding work over any backing FS that overlayfs
supports. Reflink only bounds copy-up cost. \sys uses reflink-enabled XFS, on
which stock overlayfs copy-up clones extents by reference: an edit duplicates
only the 4\,KB blocks it overwrites, and unmodified extents stay shared across
generations. On other FSes (e.g., ext4), copy-up duplicates whole files
(\autoref{sec:eval-war}).

\subsection{\agentcr: Diff-based Checkpoint/Restore}
\label{sec:design-criu}

\begin{figure}[t]
\centering
\includegraphics[width=\linewidth]{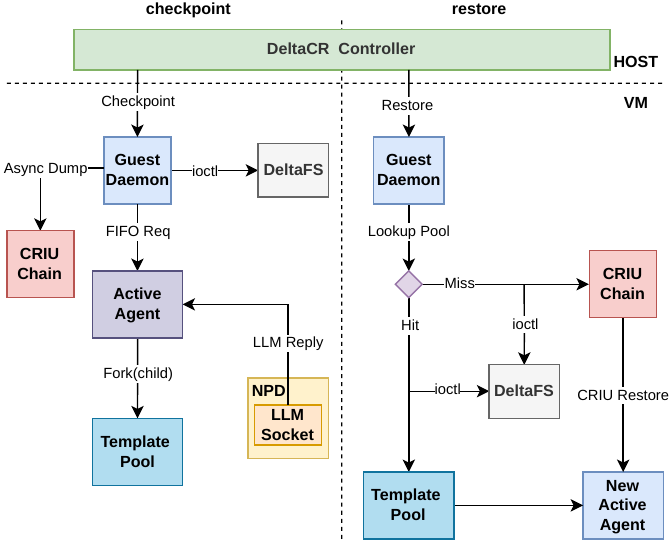}
\caption{\textbf{\agentcr architecture.} Checkpointing creates both a CRIU image and a frozen template. Restores use the template fast path on hit, or the CRIU image on miss; the network proxy (NPD) keeps external I/O off the agent path.}
\Description{\agentcr controller checkpoints an active agent into a CRIU image and a frozen template pool. Restore uses the template pool on hit, the CRIU image on miss, and a sidecar for LLM network proxying.}
\label{fig:deltacr-architecture}
\vspace{-12pt}
\end{figure}

\agentcr manages the process (memory) dimension of each checkpoint around two paths: a \emph{warm-template} fast path, which checkpoints by forking off a frozen template and restores by forking it back in low milliseconds, and a durable CRIU~\cite{criu} fallback, with incremental dumps via soft-dirty page tracking and lazy-pages restores via userfaultfd~\cite{criu-userfaultfd} (\autoref{fig:deltacr-architecture}).

\subsubsection{Template Pool and Fork-Based Restore}
\label{sec:design-template}

The fast-path restore mechanism relies on a \emph{template pool}, which maintains a registry of frozen (SIGSTOP'd) processes, each keyed to a specific snapshot ID. 

\myparagraph{Inference-masked checkpointing.}
A checkpoint runs while the agent awaits its LLM response, so both its fork and its dump are hidden inside the inference window.
Checkpoint $k$ starts once the agent finishes an action and is quiescent (no syscalls in flight, no files half-written): the agent forks, and the frozen child registers as \emph{template\_k} in the pool while the active worker continues in the main loop.
\agentcr then takes an incremental dump in the background as a durable fallback for crash recovery and cold-path restores.
Meanwhile, the agent's LLM connection lives in the network proxy (\autoref{sec:design-proxy}), so the request keeps progressing.

\myparagraph{Fast-path restore (template hit).}
To restore checkpoint $k$ while \emph{template\_k} is alive, the Guest Daemon
kills the current agent, reverts the layer stack, and forks \emph{template\_k}.
The child resumes via SIGCONT from the instruction right after the original
checkpoint, with the filesystem already switched to match.
The fork duplicates only page tables, so restore completes in low milliseconds.
An optional \hyperref[sec:design-hotcold]{async-warm thread} absorbs the
post-fork CoW faults off the critical path.
The C/R protocol makes the worker single-threaded at the moment it forks:
the fork happens only at a quiescent point, after transient thread pools are
shut down, and long-lived network threads live in the separate proxy process
(\autoref{sec:design-proxy}).
If threads or child processes remain alive at a checkpoint, \sys forgoes the
template and restores from the CRIU image, which captures the whole process tree,
including its threads, descriptors, and internal pipes.

\myparagraph{Slow-path restore (template miss).}
If \emph{template\_k} has been evicted, \agentcr falls back to CRIU lazy-pages restore.
CRIU reconstructs the process skeleton, registers all anonymous VMAs with userfaultfd, and resumes the process, faulting in pages on demand from the dump images.

\myparagraph{$N$-way fork-out (\texttt{fork\_n}).}
\label{sec:design-forkout}
RL fan-out (\autoref{sec:overview-forkout}) reuses the template-fork protocol through a batching primitive that we implement in \sys: \texttt{fork\_n} wakes the frozen template once, forks all $N$ children in one cycle, and re-suspends it, eliminating $N{-}1$ per-child resume/suspend round-trips compared with issuing $N$ one-at-a-time template restores.
FS isolation is per child: the source's overlay layers are frozen once at the branch point (one \agentfs ioctl),
and each child receives its own overlay mount whose read-only lowers are the frozen chain and whose writable upper is private, so isolating a child costs one mount and no data copy.
These mounts are pre-created before the fork, so every child resumes directly into a ready, private filesystem view.

\subsubsection{Bounding Memory under Deep Search}
\label{sec:design-mem}
\begin{figure}[t]
\centering
\includegraphics[width=\columnwidth]{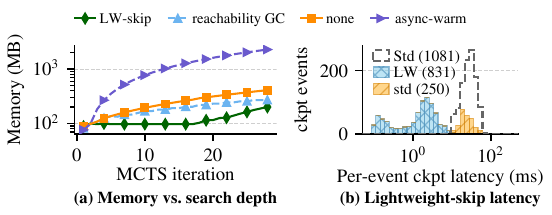}
\caption{\textbf{Bounding memory under deep search.} \textbf{(a)}~Resident memory (template pool and active worker) vs.\ MCTS iteration under four retention policies. \textbf{(b)}~Per-event checkpoint latency with and without lightweight-skip.}
\label{fig:mem}
\vspace{-12pt}
\end{figure}

By default \agentcr writes each CRIU dump to an image file on the FS, kept in the page cache for fast cold-path restore but reclaimable under memory pressure. A checkpoint's \emph{unreclaimable} resident cost is therefore just its frozen template (CoW page tables and privatized pages).
Unlike fan-out, whose children share one template's pages, deep search keeps a template per historical checkpoint, each pinning the pages its branch has since modified.
Left unbounded, resident memory grows 4.7$\times$ within 28 checkpoints (\autoref{fig:mem}a) and keeps growing until the guest exhausts memory.
\sys bounds it along three complementary axes: it \emph{avoids} creating snapshots for cheap-to-recompute nodes, \emph{caps} the resident template set, and \emph{reclaims} snapshots the search can no longer reach.

\myparagraph{Skipping read-only nodes.}
Many agent actions, such as \texttt{grep}, \texttt{cat}, and \texttt{find}, are read-only and idempotent: they leave the filesystem untouched, so re-executing one reproduces its effect on the agent.
Our rule-based classifier marks an action read-only only if every command in it, including each pipeline stage, is on a read-only allowlist (e.g., \texttt{cat}, \texttt{grep}, \texttt{sed} without \texttt{-i}) and nothing is redirected to a file. Any other or unparsable action gets a standard checkpoint.
For a read-only node, \sys skips the template fork and CRIU dump, recording only a metadata marker (parent snapshot ID and action). The layer switch still runs whenever the upper layer is dirty, so a misclassified write is not lost.
A restore to that node restores its nearest fully checkpointed ancestor and re-issues the read-only actions skipped since then, reproducing the node's state.
Because read-only steps dominate a typical trajectory (77\% of checkpoints in \autoref{fig:mem}b), this avoids creating most snapshots and sharply lowers the snapshot-store footprint, at the cost of a replay on restore.
\sys therefore enables it only when snapshot memory, rather than restore latency, is the bottleneck.

\myparagraph{Bounded template pool.}
When resident template memory becomes the bottleneck, the pool caps the live set at $N_{\text{tpl}}$ frozen templates under LRU eviction.
Evicting a template increases restore latency but preserves correctness.
Its CRIU image remains on the snapshot store, so a later re-selection falls back to the slow path (\autoref{sec:design-template}).

\myparagraph{Reachability-aware garbage collection.}
\label{sec:design-gc}
A recency- or visit-count policy is unsafe for MCTS: discarding a node UCT may still select, or an ancestor whose delta image a reachable node's incremental dump chains back to, triggers a restore-fail re-selection loop that wastes the iteration budget.
\sys instead drives GC from a persistent \emph{node-to-image index}, in which each node records its parent, children, and delta-image ID.
A pass marks the reachable set and reclaims the rest at leaf or whole-subtree granularity.
A delta image is freed only when its node is unreachable \emph{and} no reachable descendant or shared node depends on it, so reclamation never breaks the \texttt{--prev-images-dir} chain.
Since an unreachable node is never re-selected, GC drops both its frozen template and, when on-disk image storage is scarce, its delta image. GC runs asynchronously, off the checkpoint critical path.
\autoref{fig:mem}a shows the combined effect: with no policy, resident memory climbs with tree depth; lightweight-skip and reachability-GC keep it flat. Lightweight-skip additionally diverts most read-only nodes to a near-free metadata write (\autoref{fig:mem}b).

\myparagraph{Async-warm.}
\phantomsection
\label{sec:design-hotcold}
After a fork, the child shares all pages with the template using CoW, so its first write to each page takes a synchronous CoW fault.
An async-warm thread in the Guest Daemon (keeping the agent single-threaded) absorbs these faults off the critical path: right after the fork it walks the child's anonymous writable VMAs via \texttt{/proc/[pid]/mem}, forcing each page's private copy in the background while the child runs.
The latency win is not free: privatizing every page destroys CoW sharing with the template (over $5\times$ resident inflation, \autoref{fig:mem}a) and marks every touched page soft-dirty, degrading the next incremental dump to a full one.
Since post-restore CoW faults are small next to LLM and tool latency, \sys disables async-warm by default, enabling it only when they, rather than memory, are the bottleneck.

\subsubsection{Reusable I/O during Checkpoint with a Network Proxy}
\label{sec:design-proxy}

Fork-based restore is safe only if the frozen template's address space holds no long-lived external state, such as live network threads. However, Python LLM clients keep connection pools and background threads in the agent process.
Forking such a template leaves half-open sockets and locks held by threads that fork does not copy, causing runtime failures.

\sys resolves this with a \emph{network proxy daemon}, a separate process in the same sandbox that owns all LLM clients and connections.
The agent never links the SDKs. It exchanges small FIFO tokens (each $\le$\,\texttt{PIPE\_BUF}, hence atomic) with the proxy over a shared directory, keeping every socket and SDK thread out of its address space.
Being a separate process, the proxy is excluded from both the CRIU dump and the template fork, so a template carries only the agent's own short-lived threads.
During a checkpoint the agent pauses only to inline-fork. The in-flight LLM request keeps progressing inside the proxy, which buffers the response until the agent resumes.
After a fork-based restore the child inherits the FIFO descriptors from the template, resuming communication with no reconnection protocol.
Network side effects are discussed in \autoref{sec:containment}.

\subsection{\statemgr Coupling Protocol}
\label{sec:design-statemgr}

\statemgr coordinates the system to enforce an invariant: \emph{every saved state is a consistent (filesystem, memory) pair}.

\myparagraph{Consistency protocol.}
During a checkpoint, consistency rests on a \emph{quiescence invariant}:
the worker replies only after finishing its action and returning to a single-threaded control point, where it blocks on its control FIFO until the next command.
The fork (memory) and the \agentfs ioctl (filesystem) therefore observe the same post-action state, and the asynchronous dump serializes the corresponding quiescent process state.
During a restore, the restored child blocks at the same control point, and the \statemgr issues it commands only after both the layer switch and the fork have completed, so the agent never executes against mismatched files.

\myparagraph{Failure atomicity and concurrency.}
\sys is failure-atomic at the level of checkpoint visibility:
a checkpoint becomes restore-visible only at its final registration step, so a crash mid-sequence leaves the partial checkpoint unreferenced, and registered checkpoints keep their frozen layers, which are immutable.
A slow-path restore that needs the CRIU image waits for any in-flight dump to complete; the fast path reads only the frozen template.
On restore, the template fork proceeds concurrently with the layer switch (the two touch disjoint kernel state); the synchronization point is the next-command boundary above, not the fork itself.

\myparagraph{Evaluation-time test isolation.}
Evaluating a node often runs tests (e.g., SWE-Search~\cite{swe-search}),
whose side effects such as \texttt{\_\_pycache\_\_}, pytest caches, and
temporary files would permanently pollute a stateful, in-process worker.
Stateless sandboxes avoid this only by discarding the sandbox, whereas \sys
instead uses its own millisecond C/R primitive as a transaction bracket:
the \statemgr checkpoints before the test, restores once the result is
collected, and then records the result as the node's observation and
reward. This is only practical because \sys makes coupled C/R a
millisecond-scale primitive rather than a second-scale sandbox operation.

\section{Implementation and Deployment}
\label{sec:impl}

\sys is deployed inside Firecracker~\cite{firecracker} microVMs. The design is not Firecracker-specific and carries over to container- or other VM-based sandboxes.
\agentfs adds ${\sim}1.8$K LOC of C to overlayfs, all inside its loadable kernel module. It builds against stock Linux~6.8 and needs no VFS or other kernel changes.
\agentcr is ${\sim}3.8$K LOC of Python orchestrating CRIU, the template pool, the async-warm thread, and the network proxy (\autoref{sec:design-criu}).
The fork-out path (\autoref{sec:design-forkout}) is a separate ${\sim}330$-LOC module that adds the batched \texttt{fork\_n} op to the template pool's control-FIFO protocol.
CRIU dump images and overlay uppers both reside on the XFS rootfs, where reflink CoW is block-granular. Recent dump images stay in the page cache for fast restore and are reclaimed under memory pressure.

\myparagraph{Deployment model.}
\label{sec:design-compat}
\label{sec:overview-deploy}
The \statemgr's two tiers (\autoref{sec:overview}) map onto a Host--Guest split: the host-side Sandbox Controller manages VM lifecycle and workload provisioning, while the in-VM Guest Daemon owns the entire per-event C/R sequence, keeping all latency-sensitive work in a closed loop inside the guest; the KVM host is unmodified.
Branches of a task share its VM under the containment model of \autoref{sec:containment}.

\myparagraph{Integrations.}
We integrated \sys with several agent systems via out-of-tree adapters tailored to where each keeps its state.
\emph{Sandbox-resident execution workers} (e.g., SWE-agent~\cite{sweagent}): at action boundaries, coupled C/R (\autoref{sec:design-statemgr}) checkpoints the worker and binds each search node to its checkpoint (${\sim}500$ LOC).
\emph{Host-side orchestration} (e.g., LangGraph~\cite{langgraph}): LangGraph provides no sandbox, so the integrator defines the execution boundary; the framework keeps logical state on the host, while \sys keeps the sandbox's filesystem and process state (${\sim}800$ LOC).
\emph{Multithreaded agent loops that spawn tool processes} (e.g., Codex): a wrapper launches the agent under a managed process-tree root and invokes coupled C/R at action boundaries, where CRIU captures the agent's threads and live tool subprocesses; Codex's exec-server interface can also forward tool execution into \sys (${\sim}2.2$K LOC).
Porting \sys to Kata Containers took ${\sim}600$ LOC for sidecar-worker C/R integration.

\section{Evaluation}
\label{sec:eval}

\subsection{Experimental Setup}
\label{sec:eval-setup}

\myparagraph{Hardware.}
All experiments run on a four-socket server (96 physical cores), with experiment data and snapshot images placed on an NVMe SSD.
RL training fan-out (\autoref{fig:rl}b--c) additionally uses a single-node 4-GPU cluster (96\,GB each), which also hosts the Qwen3-Coder-30B inference endpoint that provides the MCTS search trajectories.
\sys (based on a Firecracker microVM) is configured with 4\,vCPUs, 8\,GB RAM, and an XFS rootfs block device plus a read-only XFS data image (target repository source and conda environment).
Unless noted, experiments use \sys's default configuration: each checkpoint takes an incremental CRIU dump plus a warm-template fork, while the lightweight-skip and async-warm optimizations (\autoref{sec:design-mem}) are optional and left disabled.

\myparagraph{Workloads.}
We draw four MCTS-trajectory archetypes from SWE-bench Verified: \textbf{Django} (fat process), \textbf{SymPy} (read-heavy exploration), \textbf{Scientific} (Astropy, Matplotlib, scikit-learn, Xarray; process-dominated), and \textbf{Tools/small repos} (pylint, requests, pytest; lightweight).
Each 30-iteration SWE-Search~\cite{swe-search} trajectory expands up to two children per node and reaches depth 15, averaging 26.4 checkpoints and 27.8 restores.

\myparagraph{Baselines.}
All baselines capture \emph{both} dimensions, since rollback determinism would otherwise break (\autoref{sec:bg-bottleneck}).
Throughout, ck/rs report the request-to-return latency of checkpoint/restore operations.
\fcdiffdm, CubeSandbox, and E2B branch at VM granularity, whereas \replaycp, \criucp, and \sys branch processes and files inside a sandbox (\autoref{sec:containment}).
\begin{itemize}[leftmargin=*,itemsep=1pt]
\item \textbf{\replaycp:} filesystem state via \texttt{shutil.copytree} of the testbed directory; process state recovered by re-executing the recorded commands.
\item \textbf{\fcdiffdm:} filesystem state via a dm-snapshot CoW layer; process state via Firecracker's incremental VM-memory snapshots.
\item \textbf{\criucp:} filesystem state via rsync hard-link snapshots; process state via incremental CRIU dumps and CRIU lazy-pages restore of the agent process.
\item \textbf{CubeSandbox}~\cite{cubesandbox}: the self-hosted v0.3.0 runtime provides filesystem state via reflink snapshots and process state via soft-dirty incremental memory snapshots.
\item \textbf{E2B\,(diff)}~\cite{e2b}: the self-hosted \texttt{e2b-dev/infra} runtime provides filesystem state via incremental rootfs snapshots and process state via incremental VM-memory snapshots.
\end{itemize}

\begin{figure}[t]
\centering
\includegraphics[width=0.95\columnwidth]{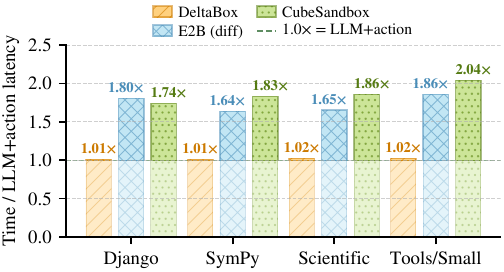}
\caption{Execution time for 30-iteration MCTS trajectories, normalized per group to each system's LLM+action latency (1.0$\times$).}
\label{fig:end2end}
\vspace{-12pt}
\end{figure}

\subsection{End-to-End Performance}
\label{sec:eval-throughput}

We evaluate two workloads that invoke coupled C/R at high frequency: MCTS search, where each tree expansion checkpoints at a parent node and restores at a node; and RL rollout fan-out, where each training step forks $N$ parallel rollout sandboxes from a single warm template.

\subsubsection{MCTS Search Throughput}
\label{sec:eval-throughput-mcts}

We evaluate MCTS performance in two stages. First, we replay recorded MCTS state-management events to measure per-event C/R latencies across the six backends in \autoref{tab:overhead}. Second, we run trajectory replays on \sys, E2B, and CubeSandbox to determine whether the per-event savings translate into lower trajectory execution time (\autoref{fig:end2end}).

\begin{table*}[t]
\caption{Per-event mean latency (ms) on SWE-bench MCTS trajectories. \emph{ck/rs} = checkpoint/restore; column naming follows \emph{process-recovery+FS-recovery} (see \autoref{sec:eval-setup} for baseline definitions).}
\label{tab:overhead}
\centering
\scriptsize
\setlength{\tabcolsep}{4pt}
\renewcommand{\arraystretch}{1.15}
\begin{tabular*}{\textwidth}{@{\extracolsep{\fill}}l rr rr rr rr rr rr}
\toprule
& \multicolumn{2}{c}{\replaycp}
& \multicolumn{2}{c}{\fcdiffdm}
& \multicolumn{2}{c}{\criucp}
& \multicolumn{2}{c}{CubeSandbox}
& \multicolumn{2}{c}{E2B\,(diff)}
& \multicolumn{2}{c}{\textbf{\sys}} \\
\cmidrule(lr){2-3}\cmidrule(lr){4-5}\cmidrule(lr){6-7}\cmidrule(lr){8-9}\cmidrule(lr){10-11}\cmidrule(lr){12-13}
\textbf{Workload} & ck$^{\dagger}$ & rs & ck & rs & ck & rs & ck & rs & ck & rs & \textbf{ck} & \textbf{rs} \\
\midrule
Django  & 631.2 & 11336 & 589.4 & 2222 & 586.7 & 975.7 & 928.3 & 758.8 & 1053 & 449.5 & \textbf{12.12} & \textbf{2.23} \\
SymPy   & 187.0 & 19887 & 767.0 & 3305 & 745.9 & 529.4 & 1008 & 638.4 & 1308 & 439.8 & \textbf{11.96} & \textbf{2.21} \\
Scientific & 303.2 & 15224 & 672.8 & 2643 & 546.5 & 600.6 & 831.4 & 645.3 & 1090 & 427.3 & \textbf{10.86} & \textbf{1.82} \\
Tools/Small repos & 201.8 & 10625 & 566.0 & 2173 & 468.8 & 442.6 & 681.4 & 571.2 & 895.8 & 394.6 & \textbf{9.16} & \textbf{1.46} \\
\midrule
Event Avg & 313.2 & 14221 & 647.9 & 2568 & 567.6 & 609.5 & 839.8 & 643.9 & 1072 & 424.7 & \textbf{10.83} & \textbf{1.86} \\
\bottomrule
\end{tabular*}

\vspace{2pt}
{\footnotesize $^{\dagger}$For \replaycp, ck is the per-trace pristine-repo copy and rs replays recorded completions. Entries are per-event means; the Event Avg is event-weighted across archetypes by each backend's own event counts.}
\end{table*}

\myparagraph{Checkpoint costs.}
\sys's standard checkpoint creates a warm template and submits an asynchronous incremental CRIU dump, both masked by LLM inference (\autoref{sec:design-criu}).
\fcdiffdm, E2B, and CubeSandbox all take incremental VM-level snapshots, yet still average 0.57--1.31\,s per checkpoint (\autoref{tab:overhead}).

\myparagraph{Restore costs.}
Restore lies directly on the agent's critical path because the next MCTS iteration cannot begin until it completes. \sys restores a checkpoint in 1.46--2.23\,ms across workloads, with an event-weighted mean of 1.86\,ms (\autoref{tab:overhead}). It achieves this low latency by overlapping the \agentfs layer switch with a fork from the warm template; template-pool management and the slow-path fallback are discussed in \autoref{sec:design-criu}.

The baselines perform substantially more work during restore. E2B and CubeSandbox restore an entire VM at every rollback, taking 0.39--0.45\,s and 0.57--0.76\,s, respectively. Most of this time goes to waiting for the agent in the resumed guest to respond (\autoref{fig:cr-cost}), a step \sys skips by restoring processes inside a running VM. FC-Diff creates a reflink copy of the base image and merges the diff chain along the MCTS ancestor path, taking 2.17--3.31\,s. Replay cold-rebuilds the target search state, including agent startup, index loading, and action re-execution, taking approximately 14.2\,s. E2B, CubeSandbox, FC-Diff, and replay are therefore approximately $230\times$, $350\times$, $1400\times$, and $7600\times$ slower than \sys, respectively.

As shown in \autoref{fig:end2end}, we replay the same 12 Qwen3-Coder-30B MCTS trajectories on \sys, self-hosted E2B\,(diff), and CubeSandbox. \sys stays within 1.01--1.02$\times$ of the LLM+action reference, whereas E2B reaches 1.64--1.86$\times$ and CubeSandbox 1.74--2.04$\times$; C/R accounts for 41.5\% and 46.3\% of their execution time, respectively. Checkpoints contribute most of this overhead (71\% for E2B and 55\% for CubeSandbox), because each VM snapshot blocks the agent for 0.7--1.3\,s (\autoref{tab:overhead}). \sys instead overlaps its incremental CRIU dump with LLM inference, leaving only millisecond-scale \agentfs switching and restore on the iteration path.

\subsubsection{RL Rollout Fan-out}
\label{sec:eval-throughput-rl}

A motivating workload for \sys is RL training, in which a single warm template spawns $N$ parallel children per training step. The underlying coupled fork-out primitive (\autoref{sec:overview-forkout}) is cheap and scales well: across 9 SWE-bench MCTS trajectories, each per-child spawn (a template fork plus the \agentfs overlay layer-switch) stays at $0.55$--$5.1$\,ms for $N$ up to 64, and each child shares the template's pages using CoW so its marginal memory cost is only the pages it later writes.

\begin{figure*}[t]
\centering
\includegraphics[width=\linewidth]{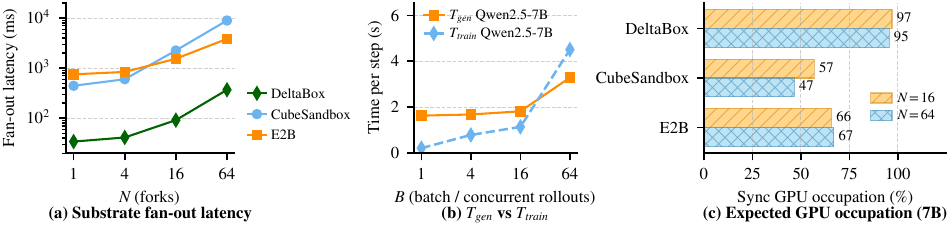}
\caption{RL training fan-out characterization. \textbf{(a)} End-to-end time to fork $N{\in}\{1,4,16,64\}$ memory-bearing children from one frozen source (64\,MiB source, each child reads it back and verifies). \textbf{(b)} $T_\text{gen}$ and $T_\text{train}$ for Qwen2.5-7B (LoRA-r16, 4 GPUs). \textbf{(c)} Expected synchronous GPU occupation $U_{\mathrm{sync}}$ at $N{\in}\{16,64\}$ (\autoref{eq:sync-gpu-occupation}).}
\label{fig:rl}
\vspace{-12pt}
\end{figure*}

\myparagraph{End-to-end sandbox fan-out.}
The preceding measurement isolates \sys's batched fork-out primitive.
We next evaluate whether this advantage carries over to complete sandbox fan-out, where each child must inherit the source process state and receive an isolated filesystem view.
\autoref{fig:rl}(a) compares the native fan-out paths of the three agent sandboxes.
Starting from a frozen source that has touched 64\,MiB of memory, each system creates $N$ children using its native fork/clone interface (\sys{} \texttt{fork\_n}, CubeSandbox \texttt{clone}, E2B \texttt{createSnapshot}$+$\texttt{create}) and is timed until every child has read back and verified the inherited state.

\sys{} is $10.4\times$--$24.7\times$ faster than both baselines (\autoref{fig:rl}(a)).

\myparagraph{Implication for synchronous RL training.}
Let $T_{\mathrm{sandbox}}$ denote the time to create all $N$ rollout sandboxes (GPU idle), $T_{\mathrm{gen}}$ the GPU rollout-generation time, and $T_{\mathrm{train}}$ the GPU training time.
Synchronous GPU occupation over a complete training step is then
\begin{equation}
U_{\mathrm{sync}}
=
\frac{T_{\mathrm{gen}} + T_{\mathrm{train}}}
     {T_{\mathrm{sandbox}} + T_{\mathrm{gen}} + T_{\mathrm{train}}}.
\label{eq:sync-gpu-occupation}
\end{equation}

We measure $T_\text{gen}$ using batched vLLM~\cite{vllm} and $T_\text{train}$ using Qwen2.5-7B LoRA forward and backward passes across $N{=}1$--$64$ rollouts (\autoref{fig:rl}(b)).
Applying \autoref{eq:sync-gpu-occupation} at $N{\in}\{16,64\}$ shows that \sys{} sustains near-saturating GPU occupation ($95$--$97$\%), while CubeSandbox reaches $47$--$57$\% and E2B reaches $66$--$67$\% (\autoref{fig:rl}(c)).

\myparagraph{Asynchronous training.}
Production frameworks such as verl~\cite{verl} (\texttt{fully\_async}) and OpenRLHF~\cite{openrlhf} decouple the trainer from the rollout workers, removing sandbox fan-out from the training critical path at the cost of rollout \emph{staleness}: a rollout is $S=(T_{\mathrm{sandbox}}+T_{\mathrm{gen}})/T_{\mathrm{train}}$ policy versions old when consumed, reusing the same three quantities as \autoref{eq:sync-gpu-occupation}.
We adopt $S{<}1$ (at most one stale version) as the target.
At $N{=}16$ the trainer step is short enough that all three backends exceed it: 1.67 for \sys{}, 3.56 for CubeSandbox, and 2.94 for E2B, with \sys{} closest.
At $N{=}64$, the larger training batch increases $T_{\mathrm{train}}$ to $4.51$\,s. \sys{}'s $S$ drops to 0.81, meeting the $S{<}1$ target, while CubeSandbox (2.71) and E2B (1.58) fail.

\subsection{Extended Studies}

\subsubsection{Checkpoint/Restore Latency}
\label{sec:eval-latency}

We present the detailed latency breakdown in \autoref{tab:latency}, alongside the corresponding phases of \criucp to separate the contributions of \agentfs and \agentcr.
\agentfs replaces the file copy with a sub-millisecond layer switch, and \agentcr replaces the synchronous dump and lazy-pages restore with an asynchronous dump and a millisecond-scale template fork.

\myparagraph{Checkpoint latency.}
Per-step checkpoint latency is dominated by the template-fork tail. The incremental CRIU dump runs asynchronously (\autoref{tab:latency}).

\myparagraph{Restore latency.}
On the fast path, the template fork dominates and the \agentfs ioctl overlaps with it inside the fork window.
The slow path, taken only after template eviction, is dominated by CRIU restore (\autoref{tab:latency}).
The layer switch likewise overlaps CRIU on the slow path. In contrast, \criucp must finish its file copy before CRIU reopens files by path, so its restore phases add up.
Both slow paths use CRIU lazy-pages restore.
\sys keeps persistent LLM connections and SDK threads in a separate proxy daemon (\autoref{sec:design-proxy}), outside the process state reconstructed by CRIU.
CRIU restore latency increases with incremental-chain length (\autoref{fig:war}(d)).
\sys implements asynchronous merging of incremental checkpoint chains to mitigate this overhead.

\begin{table}[t]
\caption{Per-component C/R latency (ms) for \criucp and \sys over the SWE-bench MCTS replay. In \sys, the fast path forks the warm template. The slow path uses the default CRIU-lazy restore mechanism.}
\label{tab:latency}
\centering
\small
\setlength{\tabcolsep}{3pt}
\begin{tabular}{@{}>{\raggedright\arraybackslash}p{0.33\columnwidth}rr@{\hspace{8pt}}rrr@{}}
\toprule
& \multicolumn{2}{c}{\criucp} & \multicolumn{3}{c}{\textbf{\sys}} \\
\cmidrule(lr){2-3}\cmidrule(lr){4-6}
\textbf{Component} & \textbf{ck} & \textbf{rs} & \textbf{ck} & \textbf{rs$_\text{fast}$} & \textbf{rs$_\text{slow}$} \\
\midrule
Filesystem                     & 143.4 & 334.0 & 0.07  & 0.19 & 0.25 \\
Fork                           & ---   & ---   & 8.87  & 1.34 & ---   \\
CRIU C/R                       & 348.8 & 102.7 & async & ---  & 8.72 \\
Coordination\textsuperscript{\S}            & 205.1 & 176.5 & ---   & 0.52 & 0.57 \\
\midrule
\textbf{Agent-perceived blocking} & $91.1^{\|}$ & 613.2 & $\mathbf{0}^{\|}$ & \textbf{1.86} & \textbf{9.29} \\
\bottomrule
\end{tabular}

\vspace{2pt}
{\footnotesize\raggedright \textsuperscript{\S} Serialized coordination outside the FS, fork, and CRIU phases.\\
$^{\|}$Mean checkpoint blocking after inference overlap: 91.1\,ms for \criucp due to unhidden checkpoint tails, and 0\,ms for \sys.\par}
\vspace{-12pt}
\end{table}

\subsubsection{Write Amplification}
\label{sec:eval-war}

\begin{figure}[t]
\centering
\includegraphics[width=0.95\columnwidth]{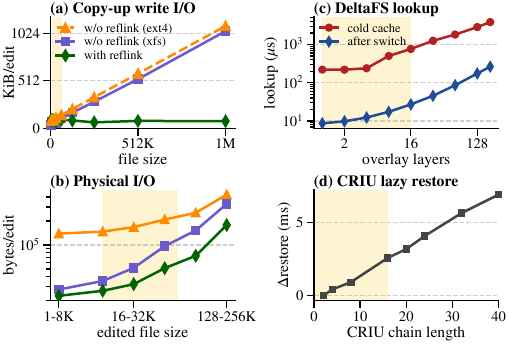}
\caption{(a)~Write I/O for fixed 4\,KiB overwrites vs.\ file size; (b)~physical I/O per replayed SWE-bench agent edit vs.\ edited-file size; (c)~\agentfs lookup latency of a bottom-layer file vs.\ layer depth; (d)~CRIU restore time added by the CRIU incremental chain. Shaded: typical edited-file sizes (a, b) and our search-tree depths (c, d).}
\label{fig:war}
\end{figure}

We measure physical I/O (bytes written to the block device, including journal and metadata) across ext4, XFS, and XFS+reflink under two workloads: a fixed 4\,KiB overwrite of 4\,KiB--1\,MiB files, which isolates the copy-up cost (\autoref{fig:war}(a)), and replayed real SWE-Search agent edits on 1--256\,KiB files (\autoref{fig:war}(b)).

\myparagraph{Copy-up savings.}
\autoref{fig:war}(a) shows that ext4 and XFS without reflink both recopy the whole file on every edit, growing linearly with file size, whereas reflink copy-up stays flat because it duplicates only the dirtied blocks.

\myparagraph{Physical I/O.}
\autoref{fig:war}(b) shows that XFS's lighter metadata dominates the I/O savings on small files (136\,KiB $\rightarrow$ 27\,KiB at 1--8\,KiB), and reflink's block sharing on large files (320.5\,KiB $\rightarrow$ 174\,KiB at 128--256\,KiB).

\subsubsection{Correctness}
\label{sec:eval-correctness}
An integration suite of 61 checks drives the \agentfs ioctl directly, covering checkpoint semantics across generations, file and directory operations after a switch, stale and deleted-but-open descriptors, mmap, and copy-up integrity; all pass.
Invalid requests are rejected atomically with the mount still writable, and layer chains stay valid up to overlayfs's 500-layer cap.

\vspace{-12pt}\section{Related Work}
\label{sec:relat}

\myparagraph{Agent sandboxes.}
Recent agent sandboxes~\cite{daytona,e2b,zeroboot,ftsandbox} primarily optimize isolation and startup latency, but provide only coarse-grained rollback.
Daytona's OCI workspaces~\cite{daytona} use Docker-layer commits and do not capture live process state.
E2B~\cite{e2b} uses Firecracker microVMs and pre-warmed templates, improving isolation but retaining VM-granularity snapshot costs. 
ZeroBoot~\cite{zeroboot} pushes VM cloning further with KVM-level CoW memory, but it forks from a fixed template whose update needs a full ${\sim}15$\,s re-snapshot without incremental checkpoints, and each branch remains an entire VM, with a shared block device, kernel-level dirty state, and no independent rollback for multiple in-VM trajectories. 
Yan's fault-tolerant sandbox~\cite{ftsandbox} moves toward transactional agent execution, but its filesystem rollback based on tool interception and userspace CoW simulation yields second-scale checkpoints and no coupled memory snapshot.

A position paper~\cite{xu2025systemsfoundationsagenticexploration} frames agentic exploration as a systems problem and, like \sys, identifies fast forking and fork-aware state management as foundational gaps; it remains a diagnostic framing, whereas \sys contributes a concrete millisecond-scale OS mechanism.
BranchFS~\cite{wang2026forkexplorecommitos} likewise targets agentic exploration, offering fork/explore/commit branch contexts with first-commit-wins semantics. Its FUSE prototype isolates only filesystem state, whereas \sys checkpoints and restores coupled filesystem and process state.

Concurrent work Crab~\cite{wu2026crabsemanticsawarecheckpointrestoreruntime} tackles the same gap from a different angle.
Crab proposes a semantics-aware approach using an eBPF-based inspector to determine checkpoint granularities.
In contrast, \sys focuses on optimizing the full OS-level C/R mechanism itself to achieve millisecond latency.
The two approaches are complementary.
DSec~\cite{huang2026deepseekelasticcomputedsec}, DeepSeek's production platform for agentic training, hosts over 380K concurrent sandboxes. Like \sys, it layers sandbox filesystems with overlayfs, but composes the layers only at creation and does not roll back running sandboxes; \sys switches layers at runtime to restore coupled filesystem and process state.

\myparagraph{Sandboxes for serverless computing.}
Serverless systems have long optimized sandbox reuse, cold start, and snapshot restore, but their unit of reuse is a function invocation rather than a rollback point inside a long-lived agent trajectory. 
TrEnv-X~\cite{trenv-x} repurposes sandboxes across invocations using OS-level memory templates backed by CXL or RDMA memory pools, while earlier cold-start systems accelerate process creation, language runtime initialization, or post-restore fault handling for short-lived functions~\cite{catalyzer,seuss,sock,reap,faasnap,spice}; Spice~\cite{spice} moves restore-path fault resolution into the kernel with Overlay VMAs. 
Dandelion~\cite{10.1145/3731569.3764803} cold-starts syscall-free compute functions in hundreds of microseconds, but runs each function to completion and keeps no long-lived execution state to roll back.
They are orthogonal to \sys. A serverless instance typically restores a compute image and then runs one request, whereas \sys repeatedly descends and backtracks within one task.

\myparagraph{Checkpointing optimizations.}
Existing checkpointing systems~\cite{10.1145/3698038.3698510,557874,10.1145/2814576.2814802,470583,10.1145/3477132.3483563,10.5555/1267411.1267429} provide important building blocks, but none provide the coupled, fine-grained rollback interface needed by stateful agents.
CRIU~\cite{criu} (which \agentcr builds on) and DMTCP~\cite{dmtcp} do process-level C/R but not the coupled, high-frequency rollback agents require. Firecracker~\cite{firecracker} and Sabre~\cite{sabre} restore whole VMs, too coarse for branches isolated inside one VM. Btrfs~\cite{btrfs} and ZFS~\cite{zfs} snapshots give block-level CoW but need a separate storage stack, and EROFS~\cite{erofs} offers read-only overlay layers without a writable upperdir; \agentfs instead hot-switches overlayfs layers over XFS reflinks. LangGraph~\cite{langgraph} and LangChain~\cite{langchain} checkpoint only logical agent state, which cannot undo executed commands' OS-level effects.

\vspace{-12pt}\section{Conclusion}
\label{sec:concl}

We present \sys, an OS-level rollbackable sandbox designed to accelerate agent workloads such as test-time tree search and RL.
Recognizing that subsequent agent states are highly similar, \sys eschews full state duplication in favor of diff-based C/R. To achieve this, \agentfs enables dynamic, unmount-free overlayfs layer switching for filesystem C/R, while \agentcr utilizes CRIU dumps and warm-template forking for memory C/R.
The \sys artifact is available at \url{https://doi.org/10.5281/zenodo.23202152}.

\begin{acks}

We are grateful to our shepherd for his detailed suggestions.
We thank the anonymous ACM ATC reviewers for their constructive feedback.
This work was supported in part by
the Fundamental and Interdisciplinary Disciplines Breakthrough Plan of the Ministry of Education of China (JYB2025XDXM113),
the National Natural Science Foundation of China (No.\,62472279 and No.\,62432010),
and the Advanced Research Center for Agent-Oriented OS (TC20260106010).
Corresponding author: Dong Du (\url{dd_nirvana@sjtu.edu.cn}).

\end{acks}


\begin{thebibliography}{68}


\ifx \showCODEN    \undefined \def \showCODEN     #1{\unskip}     \fi
\ifx \showDOI      \undefined \def \showDOI       #1{#1}\fi
\ifx \showISBNx    \undefined \def \showISBNx     #1{\unskip}     \fi
\ifx \showISBNxiii \undefined \def \showISBNxiii  #1{\unskip}     \fi
\ifx \showISSN     \undefined \def \showISSN      #1{\unskip}     \fi
\ifx \showLCCN     \undefined \def \showLCCN      #1{\unskip}     \fi
\ifx \shownote     \undefined \def \shownote      #1{#1}          \fi
\ifx \showarticletitle \undefined \def \showarticletitle #1{#1}   \fi
\ifx \showURL      \undefined \def \showURL       {\relax}        \fi
\providecommand\bibfield[2]{#2}
\providecommand\bibinfo[2]{#2}
\providecommand\natexlab[1]{#1}
\providecommand\showeprint[2][]{arXiv:#2}

\bibitem[\protect\citeauthoryear{Jimenez, Yang, Wettig, Yao, Pei, Press, and
  Narasimhan}{Jimenez et~al\mbox{.}}{2024}]{swebench}
\bibfield{author}{\bibinfo{person}{Carlos~E Jimenez}, \bibinfo{person}{John
  Yang}, \bibinfo{person}{Alexander Wettig}, \bibinfo{person}{Shunyu Yao},
  \bibinfo{person}{Kexin Pei}, \bibinfo{person}{Ofir Press}, {and}
  \bibinfo{person}{Karthik Narasimhan}.} \bibinfo{year}{2024}\natexlab{}.
\newblock \showarticletitle{SWE-bench: Can Language Models Resolve Real-world
  Github Issues?}. In \bibinfo{booktitle}{\emph{International Conference on
  Learning Representations}}, \bibfield{editor}{\bibinfo{person}{B.~Kim},
  \bibinfo{person}{Y.~Yue}, \bibinfo{person}{S.~Chaudhuri},
  \bibinfo{person}{K.~Fragkiadaki}, \bibinfo{person}{M.~Khan}, {and}
  \bibinfo{person}{Y.~Sun}} (Eds.), Vol.~\bibinfo{volume}{2024}.
  \bibinfo{pages}{54107--54157}.
\newblock
\urldef\tempurl\url{https://proceedings.iclr.cc/paper_files/paper/2024/file/edac78c3e300629acfe6cbe9ca88fb84-Paper-Conference.pdf}
\showURL{\tempurl}


\bibitem[\protect\citeauthoryear{Zhou, Xu, Zhu, Zhou, Lo, Sridhar, Cheng, Ou,
  Bisk, Fried, Alon, and Neubig}{Zhou et~al\mbox{.}}{2024}]{webarena}
\bibfield{author}{\bibinfo{person}{Shuyan Zhou}, \bibinfo{person}{Frank~F Xu},
  \bibinfo{person}{Hao Zhu}, \bibinfo{person}{Xuhui Zhou},
  \bibinfo{person}{Robert Lo}, \bibinfo{person}{Abishek Sridhar},
  \bibinfo{person}{Xianyi Cheng}, \bibinfo{person}{Tianyue Ou},
  \bibinfo{person}{Yonatan Bisk}, \bibinfo{person}{Daniel Fried},
  \bibinfo{person}{Uri Alon}, {and} \bibinfo{person}{Graham Neubig}.}
  \bibinfo{year}{2024}\natexlab{}.
\newblock \showarticletitle{WebArena: A Realistic Web Environment for Building
  Autonomous Agents}. In \bibinfo{booktitle}{\emph{International Conference on
  Learning Representations}}, \bibfield{editor}{\bibinfo{person}{B.~Kim},
  \bibinfo{person}{Y.~Yue}, \bibinfo{person}{S.~Chaudhuri},
  \bibinfo{person}{K.~Fragkiadaki}, \bibinfo{person}{M.~Khan}, {and}
  \bibinfo{person}{Y.~Sun}} (Eds.), Vol.~\bibinfo{volume}{2024}.
  \bibinfo{pages}{15585--15606}.
\newblock
\urldef\tempurl\url{https://proceedings.iclr.cc/paper_files/paper/2024/file/4410c0711e9154a7a2d26f9b3816d1ef-Paper-Conference.pdf}
\showURL{\tempurl}


\bibitem[\protect\citeauthoryear{Zhang, Feng, Zhao, Zhao, Gong, Sun, Du, Hua,
  Xia, and Chen}{Zhang et~al\mbox{.}}{2025}]{zhang2025mobiagentsystematicframeworkcustomizable}
\bibfield{author}{\bibinfo{person}{Cheng Zhang}, \bibinfo{person}{Erhu Feng},
  \bibinfo{person}{Xi Zhao}, \bibinfo{person}{Yisheng Zhao},
  \bibinfo{person}{Wangbo Gong}, \bibinfo{person}{Jiahui Sun},
  \bibinfo{person}{Dong Du}, \bibinfo{person}{Zhichao Hua},
  \bibinfo{person}{Yubin Xia}, {and} \bibinfo{person}{Haibo Chen}.}
  \bibinfo{year}{2025}\natexlab{}.
\newblock \bibinfo{title}{MobiAgent: A Systematic Framework for Customizable
  Mobile Agents}.
\newblock
\newblock
\showeprint[arxiv]{2509.00531}~[cs.MA]
\urldef\tempurl\url{https://arxiv.org/abs/2509.00531}
\showURL{\tempurl}


\bibitem[\protect\citeauthoryear{Yao, Zhao, Yu, Du, Shafran, Narasimhan, and
  Cao}{Yao et~al\mbox{.}}{2023}]{yao2023react}
\bibfield{author}{\bibinfo{person}{Shunyu Yao}, \bibinfo{person}{Jeffrey Zhao},
  \bibinfo{person}{Dian Yu}, \bibinfo{person}{Nan Du}, \bibinfo{person}{Izhak
  Shafran}, \bibinfo{person}{Karthik Narasimhan}, {and} \bibinfo{person}{Yuan
  Cao}.} \bibinfo{year}{2023}\natexlab{}.
\newblock \showarticletitle{{ReAct}: Synergizing Reasoning and Acting in
  Language Models}. In \bibinfo{booktitle}{\emph{International Conference on
  Learning Representations (ICLR)}}.
\newblock


\bibitem[\protect\citeauthoryear{{E2B}}{{E2B}}{2026}]{e2b}
\bibfield{author}{\bibinfo{person}{{E2B}}.} \bibinfo{year}{2026}\natexlab{}.
\newblock \bibinfo{title}{{E2B Documentation}}.
\newblock \bibinfo{howpublished}{\url{https://e2b.dev/docs}}.
\newblock
\newblock
\shownote{Accessed: 2026-06-10}.


\bibitem[\protect\citeauthoryear{Snell, Lee, Xu, and Kumar}{Snell
  et~al\mbox{.}}{2025}]{snell2024scaling}
\bibfield{author}{\bibinfo{person}{Charlie Snell}, \bibinfo{person}{Jaehoon
  Lee}, \bibinfo{person}{Kelvin Xu}, {and} \bibinfo{person}{Aviral Kumar}.}
  \bibinfo{year}{2025}\natexlab{}.
\newblock \showarticletitle{Scaling LLM Test-Time Compute Optimally Can be More
  Effective than Scaling Parameters for Reasoning}. In
  \bibinfo{booktitle}{\emph{International Conference on Learning
  Representations}}, \bibfield{editor}{\bibinfo{person}{Y.~Yue},
  \bibinfo{person}{A.~Garg}, \bibinfo{person}{N.~Peng},
  \bibinfo{person}{F.~Sha}, {and} \bibinfo{person}{R.~Yu}} (Eds.),
  Vol.~\bibinfo{volume}{2025}. \bibinfo{pages}{10131--10165}.
\newblock
\urldef\tempurl\url{https://proceedings.iclr.cc/paper_files/paper/2025/file/1b623663fd9b874366f3ce019fdfdd44-Paper-Conference.pdf}
\showURL{\tempurl}


\bibitem[\protect\citeauthoryear{Zhou, Yan, Shlapentokh-Rothman, Wang, and
  Wang}{Zhou et~al\mbox{.}}{2024}]{lats}
\bibfield{author}{\bibinfo{person}{Andy Zhou}, \bibinfo{person}{Kai Yan},
  \bibinfo{person}{Michal Shlapentokh-Rothman}, \bibinfo{person}{Haohan Wang},
  {and} \bibinfo{person}{Yu-Xiong Wang}.} \bibinfo{year}{2024}\natexlab{}.
\newblock \showarticletitle{Language agent tree search unifies reasoning,
  acting, and planning in language models}. In
  \bibinfo{booktitle}{\emph{Proceedings of the 41st International Conference on
  Machine Learning}} (Vienna, Austria) \emph{(\bibinfo{series}{ICML'24})}.
  \bibinfo{publisher}{JMLR.org}, Article \bibinfo{articleno}{2572},
  \bibinfo{numpages}{23}~pages.
\newblock


\bibitem[\protect\citeauthoryear{Yao, Yu, Zhao, Shafran, Griffiths, Cao, and
  Narasimhan}{Yao et~al\mbox{.}}{2023}]{NEURIPS2023_271db992}
\bibfield{author}{\bibinfo{person}{Shunyu Yao}, \bibinfo{person}{Dian Yu},
  \bibinfo{person}{Jeffrey Zhao}, \bibinfo{person}{Izhak Shafran},
  \bibinfo{person}{Tom Griffiths}, \bibinfo{person}{Yuan Cao}, {and}
  \bibinfo{person}{Karthik Narasimhan}.} \bibinfo{year}{2023}\natexlab{}.
\newblock \showarticletitle{Tree of Thoughts: Deliberate Problem Solving with
  Large Language Models}. In \bibinfo{booktitle}{\emph{Advances in Neural
  Information Processing Systems}}, \bibfield{editor}{\bibinfo{person}{A.~Oh},
  \bibinfo{person}{T.~Naumann}, \bibinfo{person}{A.~Globerson},
  \bibinfo{person}{K.~Saenko}, \bibinfo{person}{M.~Hardt}, {and}
  \bibinfo{person}{S.~Levine}} (Eds.), Vol.~\bibinfo{volume}{36}.
  \bibinfo{publisher}{Curran Associates, Inc.}, \bibinfo{pages}{11809--11822}.
\newblock
\urldef\tempurl\url{https://proceedings.neurips.cc/paper_files/paper/2023/file/271db9922b8d1f4dd7aaef84ed5ac703-Paper-Conference.pdf}
\showURL{\tempurl}


\bibitem[\protect\citeauthoryear{{The OpenClaw Project}}{{The OpenClaw
  Project}}{2026}]{openclaw}
\bibfield{author}{\bibinfo{person}{{The OpenClaw Project}}.}
  \bibinfo{year}{2026}\natexlab{}.
\newblock \bibinfo{title}{openclaw/openclaw: Your own personal AI assistant.
  Any OS. Any Platform. The lobster way.}
\newblock \bibinfo{howpublished}{\url{https://github.com/openclaw/openclaw}}.
\newblock


\bibitem[\protect\citeauthoryear{Brown, Juravsky, Ehrlich, Clark, Le, R{\'e},
  and Mirhoseini}{Brown et~al\mbox{.}}{2024}]{bon-scaling}
\bibfield{author}{\bibinfo{person}{Bradley Brown}, \bibinfo{person}{Jordan
  Juravsky}, \bibinfo{person}{Ryan Ehrlich}, \bibinfo{person}{Ronald Clark},
  \bibinfo{person}{Quoc~V. Le}, \bibinfo{person}{Christopher R{\'e}}, {and}
  \bibinfo{person}{Azalia Mirhoseini}.} \bibinfo{year}{2024}\natexlab{}.
\newblock \bibinfo{title}{Large Language Monkeys: Scaling Inference Compute
  with Repeated Sampling}.
\newblock
\newblock
\showeprint[arxiv]{2407.21787}~[cs.LG]
\urldef\tempurl\url{https://arxiv.org/abs/2407.21787}
\showURL{\tempurl}


\bibitem[\protect\citeauthoryear{He, Li, Feng, Du, Liu, Liu, Xia, and Chen}{He
  et~al\mbox{.}}{2026}]{10.1145/3779212.3790172}
\bibfield{author}{\bibinfo{person}{Jingkai He}, \bibinfo{person}{Tianjian Li},
  \bibinfo{person}{Erhu Feng}, \bibinfo{person}{Dong Du}, \bibinfo{person}{Qian
  Liu}, \bibinfo{person}{Tao Liu}, \bibinfo{person}{Yubin Xia}, {and}
  \bibinfo{person}{Haibo Chen}.} \bibinfo{year}{2026}\natexlab{}.
\newblock \showarticletitle{History Doesn't Repeat Itself but Rollouts Rhyme:
  Accelerating Reinforcement Learning with RhymeRL}. In
  \bibinfo{booktitle}{\emph{Proceedings of the 31st ACM International
  Conference on Architectural Support for Programming Languages and Operating
  Systems, Volume 2}} (USA) \emph{(\bibinfo{series}{ASPLOS '26})}.
  \bibinfo{publisher}{Association for Computing Machinery},
  \bibinfo{address}{New York, NY, USA}, \bibinfo{pages}{929–945}.
\newblock
\showISBNx{9798400723599}
\urldef\tempurl\url{https://doi.org/10.1145/3779212.3790172}
\showDOI{\tempurl}


\bibitem[\protect\citeauthoryear{Shao, Wang, Zhu, Xu, Song, Bi, Zhang, Zhang,
  Li, Wu, and Guo}{Shao et~al\mbox{.}}{2024}]{shao2024deepseekmathpushinglimitsmathematical}
\bibfield{author}{\bibinfo{person}{Zhihong Shao}, \bibinfo{person}{Peiyi Wang},
  \bibinfo{person}{Qihao Zhu}, \bibinfo{person}{Runxin Xu},
  \bibinfo{person}{Junxiao Song}, \bibinfo{person}{Xiao Bi},
  \bibinfo{person}{Haowei Zhang}, \bibinfo{person}{Mingchuan Zhang},
  \bibinfo{person}{Y.~K. Li}, \bibinfo{person}{Y. Wu}, {and}
  \bibinfo{person}{Daya Guo}.} \bibinfo{year}{2024}\natexlab{}.
\newblock \bibinfo{title}{DeepSeekMath: Pushing the Limits of Mathematical
  Reasoning in Open Language Models}.
\newblock
\newblock
\showeprint[arxiv]{2402.03300}~[cs.CL]
\urldef\tempurl\url{https://arxiv.org/abs/2402.03300}
\showURL{\tempurl}


\bibitem[\protect\citeauthoryear{Yu, Zhang, Zhu, Yuan, Zuo, Yue, Dai, Fan, Liu,
  Liu, Liu, Liu, Lin, Lin, Ma, Sheng, Tong, Zhang, Zhang, Zhang, Zhang, Zhu,
  Zhu, Chen, Chen, Wang, Yu, Song, Wei, Zhou, Liu, Ma, Zhang, Yan, Wu, and
  Wang}{Yu et~al\mbox{.}}{2025}]{NEURIPS2025_a4277440}
\bibfield{author}{\bibinfo{person}{Qiying Yu}, \bibinfo{person}{Zheng Zhang},
  \bibinfo{person}{Ruofei Zhu}, \bibinfo{person}{Yufeng Yuan},
  \bibinfo{person}{Xiaochen Zuo}, \bibinfo{person}{Yu Yue},
  \bibinfo{person}{Weinan Dai}, \bibinfo{person}{Tiantian Fan},
  \bibinfo{person}{Gaohong Liu}, \bibinfo{person}{Juncai Liu},
  \bibinfo{person}{Lingjun Liu}, \bibinfo{person}{Xin Liu},
  \bibinfo{person}{Haibin Lin}, \bibinfo{person}{Zhiqi Lin},
  \bibinfo{person}{Bole Ma}, \bibinfo{person}{Guangming Sheng},
  \bibinfo{person}{Yuxuan Tong}, \bibinfo{person}{Chi Zhang},
  \bibinfo{person}{Mofan Zhang}, \bibinfo{person}{Ru Zhang},
  \bibinfo{person}{Wang Zhang}, \bibinfo{person}{Hang Zhu},
  \bibinfo{person}{Jinhua Zhu}, \bibinfo{person}{Jiaze Chen},
  \bibinfo{person}{Jiangjie Chen}, \bibinfo{person}{Chengyi Wang},
  \bibinfo{person}{Hongli Yu}, \bibinfo{person}{Yuxuan Song},
  \bibinfo{person}{Xiangpeng Wei}, \bibinfo{person}{Hao Zhou},
  \bibinfo{person}{Jingjing Liu}, \bibinfo{person}{Wei-Ying Ma},
  \bibinfo{person}{Ya-Qin Zhang}, \bibinfo{person}{Lin Yan},
  \bibinfo{person}{Yonghui Wu}, {and} \bibinfo{person}{Mingxuan Wang}.}
  \bibinfo{year}{2025}\natexlab{}.
\newblock \showarticletitle{DAPO: An Open-Source LLM Reinforcement Learning
  System at Scale}. In \bibinfo{booktitle}{\emph{Advances in Neural Information
  Processing Systems}}, \bibfield{editor}{\bibinfo{person}{D.~Belgrave},
  \bibinfo{person}{C.~Zhang}, \bibinfo{person}{H.~Lin},
  \bibinfo{person}{R.~Pascanu}, \bibinfo{person}{P.~Koniusz},
  \bibinfo{person}{M.~Ghassemi}, {and} \bibinfo{person}{N.~Chen}} (Eds.),
  Vol.~\bibinfo{volume}{38}. \bibinfo{publisher}{Curran Associates, Inc.},
  \bibinfo{pages}{113222--113244}.
\newblock
\urldef\tempurl\url{https://proceedings.neurips.cc/paper_files/paper/2025/file/a4277440d50f1f15d2cb4c14f7e0c0d2-Paper-Conference.pdf}
\showURL{\tempurl}


\bibitem[\protect\citeauthoryear{{Daytona}}{{Daytona}}{2024}]{daytona}
\bibfield{author}{\bibinfo{person}{{Daytona}}.}
  \bibinfo{year}{2024}\natexlab{}.
\newblock \bibinfo{title}{{Daytona}}.
\newblock \bibinfo{howpublished}{\url{https://daytona.io}}.
\newblock


\bibitem[\protect\citeauthoryear{Yan}{Yan}{2025}]{ftsandbox}
\bibfield{author}{\bibinfo{person}{Boyang Yan}.}
  \bibinfo{year}{2025}\natexlab{}.
\newblock \bibinfo{title}{Fault-Tolerant Sandboxing for {AI} Coding Agents: A
  Transactional Approach to Safe Autonomous Execution}.
\newblock
\newblock
\showeprint[arxiv]{2512.12806}~[cs.AI]
\urldef\tempurl\url{https://arxiv.org/abs/2512.12806}
\showURL{\tempurl}


\bibitem[\protect\citeauthoryear{{The {CRIU} Project}}{{The {CRIU}
  Project}}{2011}]{criu}
\bibfield{author}{\bibinfo{person}{{The {CRIU} Project}}.}
  \bibinfo{year}{2011}\natexlab{}.
\newblock \bibinfo{title}{{CRIU}: Checkpoint/Restore In Userspace}.
\newblock \bibinfo{howpublished}{\url{https://criu.org}}.
\newblock


\bibitem[\protect\citeauthoryear{{E2B}}{{E2B}}{2026}]{e2b-checkpoint}
\bibfield{author}{\bibinfo{person}{{E2B}}.} \bibinfo{year}{2026}\natexlab{}.
\newblock \bibinfo{title}{E2B Sandbox persistence}.
\newblock
  \bibinfo{howpublished}{\url{https://e2b.dev/docs/sandbox/persistence}}.
\newblock


\bibitem[\protect\citeauthoryear{Agache, Brooker, Florescu, Iordache, Liguori,
  Neugebauer, Piwonka, and Popa}{Agache et~al\mbox{.}}{2020}]{firecracker}
\bibfield{author}{\bibinfo{person}{Alexandru Agache}, \bibinfo{person}{Marc
  Brooker}, \bibinfo{person}{Andreea Florescu}, \bibinfo{person}{Alexandra
  Iordache}, \bibinfo{person}{Anthony Liguori}, \bibinfo{person}{Rolf
  Neugebauer}, \bibinfo{person}{Phil Piwonka}, {and}
  \bibinfo{person}{Diana-Maria Popa}.} \bibinfo{year}{2020}\natexlab{}.
\newblock \showarticletitle{Firecracker: lightweight virtualization for
  serverless applications}. In \bibinfo{booktitle}{\emph{Proceedings of the
  17th Usenix Conference on Networked Systems Design and Implementation}}
  (Santa Clara, CA, USA) \emph{(\bibinfo{series}{NSDI'20})}.
  \bibinfo{publisher}{USENIX Association}, \bibinfo{address}{USA},
  \bibinfo{pages}{419--434}.
\newblock
\showISBNx{9781939133137}


\bibitem[\protect\citeauthoryear{Ao, Porter, and Voelker}{Ao
  et~al\mbox{.}}{2022}]{faasnap}
\bibfield{author}{\bibinfo{person}{Lixiang Ao}, \bibinfo{person}{George
  Porter}, {and} \bibinfo{person}{Geoffrey~M. Voelker}.}
  \bibinfo{year}{2022}\natexlab{}.
\newblock \showarticletitle{FaaSnap: FaaS made fast using snapshot-based VMs}.
  In \bibinfo{booktitle}{\emph{Proceedings of the Seventeenth European
  Conference on Computer Systems}} (Rennes, France)
  \emph{(\bibinfo{series}{EuroSys '22})}. \bibinfo{publisher}{Association for
  Computing Machinery}, \bibinfo{address}{New York, NY, USA},
  \bibinfo{pages}{730--746}.
\newblock
\showISBNx{9781450391627}
\urldef\tempurl\url{https://doi.org/10.1145/3492321.3524270}
\showDOI{\tempurl}


\bibitem[\protect\citeauthoryear{Du, Yu, Xia, Zang, Yan, Qin, Wu, and Chen}{Du
  et~al\mbox{.}}{2020}]{catalyzer}
\bibfield{author}{\bibinfo{person}{Dong Du}, \bibinfo{person}{Tianyi Yu},
  \bibinfo{person}{Yubin Xia}, \bibinfo{person}{Binyu Zang},
  \bibinfo{person}{Guanglu Yan}, \bibinfo{person}{Chenggang Qin},
  \bibinfo{person}{Qixuan Wu}, {and} \bibinfo{person}{Haibo Chen}.}
  \bibinfo{year}{2020}\natexlab{}.
\newblock \showarticletitle{Catalyzer: Sub-millisecond Startup for Serverless
  Computing with Initialization-less Booting}. In
  \bibinfo{booktitle}{\emph{Proceedings of the Twenty-Fifth International
  Conference on Architectural Support for Programming Languages and Operating
  Systems}} (Lausanne, Switzerland) \emph{(\bibinfo{series}{ASPLOS '20})}.
  \bibinfo{publisher}{Association for Computing Machinery},
  \bibinfo{address}{New York, NY, USA}, \bibinfo{pages}{467--481}.
\newblock
\showISBNx{9781450371025}
\urldef\tempurl\url{https://doi.org/10.1145/3373376.3378512}
\showDOI{\tempurl}


\bibitem[\protect\citeauthoryear{Ustiugov, Petrov, Kogias, Bugnion, and
  Grot}{Ustiugov et~al\mbox{.}}{2021}]{reap}
\bibfield{author}{\bibinfo{person}{Dmitrii Ustiugov}, \bibinfo{person}{Plamen
  Petrov}, \bibinfo{person}{Marios Kogias}, \bibinfo{person}{Edouard Bugnion},
  {and} \bibinfo{person}{Boris Grot}.} \bibinfo{year}{2021}\natexlab{}.
\newblock \showarticletitle{Benchmarking, analysis, and optimization of
  serverless function snapshots}. In \bibinfo{booktitle}{\emph{Proceedings of
  the 26th ACM International Conference on Architectural Support for
  Programming Languages and Operating Systems}} (Virtual, USA)
  \emph{(\bibinfo{series}{ASPLOS '21})}. \bibinfo{publisher}{Association for
  Computing Machinery}, \bibinfo{address}{New York, NY, USA},
  \bibinfo{pages}{559--572}.
\newblock
\showISBNx{9781450383172}
\urldef\tempurl\url{https://doi.org/10.1145/3445814.3446714}
\showDOI{\tempurl}


\bibitem[\protect\citeauthoryear{Chai, Zhou, Hu, Tan, Bie, Shen, Shen, Xing,
  Song, Yang, Gao, Yu, He, Du, Xia, Chen, and Chen}{Chai et~al\mbox{.}}{2025}]{10.5555/3767901.3767929}
\bibfield{author}{\bibinfo{person}{Xiaohu Chai}, \bibinfo{person}{Tianyu Zhou},
  \bibinfo{person}{Keyang Hu}, \bibinfo{person}{Jianfeng Tan},
  \bibinfo{person}{Tiwei Bie}, \bibinfo{person}{Anqi Shen},
  \bibinfo{person}{Dawei Shen}, \bibinfo{person}{Qi Xing},
  \bibinfo{person}{Shun Song}, \bibinfo{person}{Tongkai Yang},
  \bibinfo{person}{Le Gao}, \bibinfo{person}{Feng Yu}, \bibinfo{person}{Zhengyu
  He}, \bibinfo{person}{Dong Du}, \bibinfo{person}{Yubin Xia},
  \bibinfo{person}{Kang Chen}, {and} \bibinfo{person}{Yu Chen}.}
  \bibinfo{year}{2025}\natexlab{}.
\newblock \showarticletitle{Fork in the road: reflections and optimizations for
  cold start latency in production serverless systems}. In
  \bibinfo{booktitle}{\emph{Proceedings of the 19th USENIX Conference on
  Operating Systems Design and Implementation}} (Boston, MA, USA)
  \emph{(\bibinfo{series}{OSDI '25})}. \bibinfo{publisher}{USENIX Association},
  \bibinfo{address}{USA}, Article \bibinfo{articleno}{28},
  \bibinfo{numpages}{20}~pages.
\newblock
\showISBNx{978-1-939133-47-2}


\bibitem[\protect\citeauthoryear{Brown}{Brown}{2026}]{overlayfs}
\bibfield{author}{\bibinfo{person}{Neil Brown}.}
  \bibinfo{year}{2026}\natexlab{}.
\newblock \bibinfo{title}{Overlay Filesystem}.
\newblock
  \bibinfo{howpublished}{\url{https://docs.kernel.org/filesystems/overlayfs.html}}.
\newblock


\bibitem[\protect\citeauthoryear{Cvetkovi\'{c}, Costa, Djokic, Friedman, and
  Klimovic}{Cvetkovi\'{c} et~al\mbox{.}}{2024}]{10.1145/3694715.3695966}
\bibfield{author}{\bibinfo{person}{Lazar Cvetkovi\'{c}},
  \bibinfo{person}{Fran\c{c}ois Costa}, \bibinfo{person}{Mihajlo Djokic},
  \bibinfo{person}{Michal Friedman}, {and} \bibinfo{person}{Ana Klimovic}.}
  \bibinfo{year}{2024}\natexlab{}.
\newblock \showarticletitle{Dirigent: Lightweight Serverless Orchestration}. In
  \bibinfo{booktitle}{\emph{Proceedings of the ACM SIGOPS 30th Symposium on
  Operating Systems Principles}} (Austin, TX, USA) \emph{(\bibinfo{series}{SOSP
  '24})}. \bibinfo{publisher}{Association for Computing Machinery},
  \bibinfo{address}{New York, NY, USA}, \bibinfo{pages}{369–384}.
\newblock
\showISBNx{9798400712517}
\urldef\tempurl\url{https://doi.org/10.1145/3694715.3695966}
\showDOI{\tempurl}


\bibitem[\protect\citeauthoryear{Du, Liu, Jiang, Xia, Zang, and Chen}{Du
  et~al\mbox{.}}{2022}]{10.1145/3503222.3507732}
\bibfield{author}{\bibinfo{person}{Dong Du}, \bibinfo{person}{Qingyuan Liu},
  \bibinfo{person}{Xueqiang Jiang}, \bibinfo{person}{Yubin Xia},
  \bibinfo{person}{Binyu Zang}, {and} \bibinfo{person}{Haibo Chen}.}
  \bibinfo{year}{2022}\natexlab{}.
\newblock \showarticletitle{Serverless computing on heterogeneous computers}.
  In \bibinfo{booktitle}{\emph{Proceedings of the 27th ACM International
  Conference on Architectural Support for Programming Languages and Operating
  Systems}} (Lausanne, Switzerland) \emph{(\bibinfo{series}{ASPLOS '22})}.
  \bibinfo{publisher}{Association for Computing Machinery},
  \bibinfo{address}{New York, NY, USA}, \bibinfo{pages}{797–813}.
\newblock
\showISBNx{9781450392051}
\urldef\tempurl\url{https://doi.org/10.1145/3503222.3507732}
\showDOI{\tempurl}


\bibitem[\protect\citeauthoryear{Li, Cheng, Chen, Guan, Bian, Tao, Zha, Wang,
  Han, and Guo}{Li et~al\mbox{.}}{2022}]{280716}
\bibfield{author}{\bibinfo{person}{Zijun Li}, \bibinfo{person}{Jiagan Cheng},
  \bibinfo{person}{Quan Chen}, \bibinfo{person}{Eryu Guan},
  \bibinfo{person}{Zizheng Bian}, \bibinfo{person}{Yi Tao},
  \bibinfo{person}{Bin Zha}, \bibinfo{person}{Qiang Wang},
  \bibinfo{person}{Weidong Han}, {and} \bibinfo{person}{Minyi Guo}.}
  \bibinfo{year}{2022}\natexlab{}.
\newblock \showarticletitle{{RunD}: A Lightweight Secure Container Runtime for
  High-density Deployment and High-concurrency Startup in Serverless
  Computing}. In \bibinfo{booktitle}{\emph{2022 USENIX Annual Technical
  Conference (USENIX ATC 22)}}. \bibinfo{publisher}{USENIX Association},
  \bibinfo{address}{Carlsbad, CA}, \bibinfo{pages}{53--68}.
\newblock
\showISBNx{978-1-939133-29-27}
\urldef\tempurl\url{https://www.usenix.org/conference/atc22/presentation/li-zijun-rund}
\showURL{\tempurl}


\bibitem[\protect\citeauthoryear{Yu, Basu~Roy, Fontenot, Tiwari, Li, Zhang,
  Wang, and Park}{Yu et~al\mbox{.}}{2024}]{10.1145/3617232.3624871}
\bibfield{author}{\bibinfo{person}{Hanfei Yu}, \bibinfo{person}{Rohan
  Basu~Roy}, \bibinfo{person}{Christian Fontenot}, \bibinfo{person}{Devesh
  Tiwari}, \bibinfo{person}{Jian Li}, \bibinfo{person}{Hong Zhang},
  \bibinfo{person}{Hao Wang}, {and} \bibinfo{person}{Seung-Jong Park}.}
  \bibinfo{year}{2024}\natexlab{}.
\newblock \showarticletitle{RainbowCake: Mitigating Cold-starts in Serverless
  with Layer-wise Container Caching and Sharing}. In
  \bibinfo{booktitle}{\emph{Proceedings of the 29th ACM International
  Conference on Architectural Support for Programming Languages and Operating
  Systems, Volume 1}} (La Jolla, CA, USA) \emph{(\bibinfo{series}{ASPLOS
  '24})}. \bibinfo{publisher}{Association for Computing Machinery},
  \bibinfo{address}{New York, NY, USA}, \bibinfo{pages}{335–350}.
\newblock
\showISBNx{9798400703720}
\urldef\tempurl\url{https://doi.org/10.1145/3617232.3624871}
\showDOI{\tempurl}


\bibitem[\protect\citeauthoryear{Huang, Ma, Liu, Lin, Chen, Jiang, Liao, Shan,
  Wu, Zhang, Lu, Ma, Gong, and Zhang}{Huang et~al\mbox{.}}{2026}]{trenv-x}
\bibfield{author}{\bibinfo{person}{Jialiang Huang}, \bibinfo{person}{Teng Ma},
  \bibinfo{person}{Zheng Liu}, \bibinfo{person}{Sixing Lin},
  \bibinfo{person}{Kang Chen}, \bibinfo{person}{Jinlei Jiang},
  \bibinfo{person}{Xia Liao}, \bibinfo{person}{Yingdi Shan},
  \bibinfo{person}{Yongwei Wu}, \bibinfo{person}{Ning Zhang},
  \bibinfo{person}{Mengting Lu}, \bibinfo{person}{Tao Ma},
  \bibinfo{person}{Haifeng Gong}, {and} \bibinfo{person}{Mingxing Zhang}.}
  \bibinfo{year}{2026}\natexlab{}.
\newblock \showarticletitle{{TrEnv-X}: Transparently Share Serverless Execution
  Environments Across Different Functions and Nodes}.
\newblock \bibinfo{journal}{\emph{ACM Trans. Comput. Syst.}}
  \bibinfo{volume}{44}, \bibinfo{number}{3}, Article \bibinfo{articleno}{16}
  (\bibinfo{date}{July} \bibinfo{year}{2026}), \bibinfo{numpages}{39}~pages.
\newblock
\showISSN{0734-2071}
\urldef\tempurl\url{https://doi.org/10.1145/3805475}
\showDOI{\tempurl}


\bibitem[\protect\citeauthoryear{Szekely, Belay, Morris, and Kaashoek}{Szekely
  et~al\mbox{.}}{2024}]{10.1145/3694715.3695947}
\bibfield{author}{\bibinfo{person}{Ariel Szekely}, \bibinfo{person}{Adam
  Belay}, \bibinfo{person}{Robert Morris}, {and} \bibinfo{person}{M.~Frans
  Kaashoek}.} \bibinfo{year}{2024}\natexlab{}.
\newblock \showarticletitle{Unifying serverless and microservice workloads with
  SigmaOS}. In \bibinfo{booktitle}{\emph{Proceedings of the ACM SIGOPS 30th
  Symposium on Operating Systems Principles}} (Austin, TX, USA)
  \emph{(\bibinfo{series}{SOSP '24})}. \bibinfo{publisher}{Association for
  Computing Machinery}, \bibinfo{address}{New York, NY, USA},
  \bibinfo{pages}{385–402}.
\newblock
\showISBNx{9798400712517}
\urldef\tempurl\url{https://doi.org/10.1145/3694715.3695947}
\showDOI{\tempurl}


\bibitem[\protect\citeauthoryear{{DeepSeek-AI} et~al\mbox{.}}{{DeepSeek-AI}
  et~al\mbox{.}}{2026}]{deepseek-v4}
\bibfield{author}{\bibinfo{person}{{DeepSeek-AI}} {et~al\mbox{.}}}
  \bibinfo{year}{2026}\natexlab{}.
\newblock \bibinfo{title}{{DeepSeek-V4}: Towards Highly Efficient Million-Token
  Context Intelligence}.
\newblock
\newblock
\showeprint[arxiv]{2606.19348}~[cs.CL]
\urldef\tempurl\url{https://arxiv.org/abs/2606.19348}
\showURL{\tempurl}


\bibitem[\protect\citeauthoryear{{Tencent Cloud}}{{Tencent Cloud}}{2026}]{cubesandbox}
\bibfield{author}{\bibinfo{person}{{Tencent Cloud}}.}
  \bibinfo{year}{2026}\natexlab{}.
\newblock \bibinfo{title}{{CubeSandbox} v0.3.0: Instant, Concurrent, Secure \&
  Lightweight Sandbox for AI Agents}.
\newblock
  \bibinfo{howpublished}{\url{https://github.com/TencentCloud/CubeSandbox/releases/tag/v0.3.0}}.
\newblock
\newblock
\shownote{CubeCoW Copy-on-Write snapshot engine (reflink volume + soft-dirty
  memory); accessed June 2026}.


\bibitem[\protect\citeauthoryear{Kocsis and Szepesv\'{a}ri}{Kocsis and
  Szepesv\'{a}ri}{2006}]{kocsis2006uct}
\bibfield{author}{\bibinfo{person}{Levente Kocsis} {and} \bibinfo{person}{Csaba
  Szepesv\'{a}ri}.} \bibinfo{year}{2006}\natexlab{}.
\newblock \showarticletitle{Bandit Based Monte-Carlo Planning}. In
  \bibinfo{booktitle}{\emph{Proceedings of the 17th European Conference on
  Machine Learning (ECML)}}. \bibinfo{pages}{282--293}.
\newblock


\bibitem[\protect\citeauthoryear{Browne, Powley, Whitehouse, Lucas, Cowling,
  Rohlfshagen, Tavener, Perez, Samothrakis, and Colton}{Browne
  et~al\mbox{.}}{2012}]{browne2012mctssurvey}
\bibfield{author}{\bibinfo{person}{Cameron~B. Browne}, \bibinfo{person}{Edward
  Powley}, \bibinfo{person}{Daniel Whitehouse}, \bibinfo{person}{Simon~M.
  Lucas}, \bibinfo{person}{Peter~I. Cowling}, \bibinfo{person}{Philipp
  Rohlfshagen}, \bibinfo{person}{Stephen Tavener}, \bibinfo{person}{Diego
  Perez}, \bibinfo{person}{Spyridon Samothrakis}, {and} \bibinfo{person}{Simon
  Colton}.} \bibinfo{year}{2012}\natexlab{}.
\newblock \showarticletitle{A Survey of Monte Carlo Tree Search Methods}.
\newblock \bibinfo{journal}{\emph{IEEE Transactions on Computational
  Intelligence and AI in Games}} \bibinfo{volume}{4}, \bibinfo{number}{1}
  (\bibinfo{year}{2012}), \bibinfo{pages}{1--43}.
\newblock


\bibitem[\protect\citeauthoryear{Antoniades, \"{O}rwall, Zhang, Xie, Goyal, and
  Wang}{Antoniades et~al\mbox{.}}{2025}]{swe-search}
\bibfield{author}{\bibinfo{person}{Antonis Antoniades}, \bibinfo{person}{Albert
  \"{O}rwall}, \bibinfo{person}{Kexun Zhang}, \bibinfo{person}{Yuxi Xie},
  \bibinfo{person}{Anirudh Goyal}, {and} \bibinfo{person}{William Wang}.}
  \bibinfo{year}{2025}\natexlab{}.
\newblock \showarticletitle{SWE-Search: Enhancing Software Agents with Monte
  Carlo Tree Search and Iterative Refinement}. In
  \bibinfo{booktitle}{\emph{International Conference on Learning
  Representations}}, \bibfield{editor}{\bibinfo{person}{Y.~Yue},
  \bibinfo{person}{A.~Garg}, \bibinfo{person}{N.~Peng},
  \bibinfo{person}{F.~Sha}, {and} \bibinfo{person}{R.~Yu}} (Eds.),
  Vol.~\bibinfo{volume}{2025}. \bibinfo{pages}{64485--64515}.
\newblock
\urldef\tempurl\url{https://proceedings.iclr.cc/paper_files/paper/2025/file/a1e6783e4d739196cad3336f12d402bf-Paper-Conference.pdf}
\showURL{\tempurl}


\bibitem[\protect\citeauthoryear{Ji, Ma, Wang, Chen, Chu, and Wu}{Ji
  et~al\mbox{.}}{2026}]{ji2026tree}
\bibfield{author}{\bibinfo{person}{Yuxiang Ji}, \bibinfo{person}{Ziyu Ma},
  \bibinfo{person}{Yong Wang}, \bibinfo{person}{Guanhua Chen},
  \bibinfo{person}{Xiangxiang Chu}, {and} \bibinfo{person}{Liaoni Wu}.}
  \bibinfo{year}{2026}\natexlab{}.
\newblock \showarticletitle{Tree Search for {LLM} Agent Reinforcement
  Learning}. In \bibinfo{booktitle}{\emph{The Fourteenth International
  Conference on Learning Representations}}.
\newblock
\urldef\tempurl\url{https://openreview.net/forum?id=ZpQwAFhU13}
\showURL{\tempurl}


\bibitem[\protect\citeauthoryear{Yang, Jimenez, Wettig, Lieret, Yao,
  Narasimhan, and Press}{Yang et~al\mbox{.}}{2024}]{sweagent}
\bibfield{author}{\bibinfo{person}{John Yang}, \bibinfo{person}{Carlos
  Jimenez}, \bibinfo{person}{Alexander Wettig}, \bibinfo{person}{Kilian
  Lieret}, \bibinfo{person}{Shunyu Yao}, \bibinfo{person}{Karthik Narasimhan},
  {and} \bibinfo{person}{Ofir Press}.} \bibinfo{year}{2024}\natexlab{}.
\newblock \showarticletitle{SWE-agent: Agent-Computer Interfaces Enable
  Automated Software Engineering}. In \bibinfo{booktitle}{\emph{Advances in
  Neural Information Processing Systems}},
  \bibfield{editor}{\bibinfo{person}{A.~Globerson},
  \bibinfo{person}{L.~Mackey}, \bibinfo{person}{D.~Belgrave},
  \bibinfo{person}{A.~Fan}, \bibinfo{person}{U.~Paquet},
  \bibinfo{person}{J.~Tomczak}, {and} \bibinfo{person}{C.~Zhang}} (Eds.),
  Vol.~\bibinfo{volume}{37}. \bibinfo{publisher}{Curran Associates, Inc.},
  \bibinfo{pages}{50528--50652}.
\newblock
\urldef\tempurl\url{https://doi.org/10.52202/079017-1601}
\showDOI{\tempurl}


\bibitem[\protect\citeauthoryear{Wang, Li, Song, Xu, Tang, Zhuge, Pan, Song,
  Li, Singh, Tran, Li, Ma, Zheng, Qian, Shao, Muennighoff, Zhang, Hui, Lin,
  Brennan, Peng, Ji, and Neubig}{Wang et~al\mbox{.}}{2025}]{openhands}
\bibfield{author}{\bibinfo{person}{Xingyao Wang}, \bibinfo{person}{Boxuan Li},
  \bibinfo{person}{Yufan Song}, \bibinfo{person}{Frank~F Xu},
  \bibinfo{person}{Xiangru Tang}, \bibinfo{person}{Mingchen Zhuge},
  \bibinfo{person}{Jiayi Pan}, \bibinfo{person}{Yueqi Song},
  \bibinfo{person}{Bowen Li}, \bibinfo{person}{Jaskirat Singh},
  \bibinfo{person}{Hoang Tran}, \bibinfo{person}{Fuqiang Li},
  \bibinfo{person}{Ren Ma}, \bibinfo{person}{Mingzhang Zheng},
  \bibinfo{person}{Bill Qian}, \bibinfo{person}{Daniel Shao},
  \bibinfo{person}{Niklas Muennighoff}, \bibinfo{person}{Yizhe Zhang},
  \bibinfo{person}{Binyuan Hui}, \bibinfo{person}{Junyang Lin},
  \bibinfo{person}{Robert Brennan}, \bibinfo{person}{Hao Peng},
  \bibinfo{person}{Heng Ji}, {and} \bibinfo{person}{Graham Neubig}.}
  \bibinfo{year}{2025}\natexlab{}.
\newblock \showarticletitle{OpenHands: An Open Platform for AI Software
  Developers as Generalist Agents}. In \bibinfo{booktitle}{\emph{International
  Conference on Learning Representations}},
  \bibfield{editor}{\bibinfo{person}{Y.~Yue}, \bibinfo{person}{A.~Garg},
  \bibinfo{person}{N.~Peng}, \bibinfo{person}{F.~Sha}, {and}
  \bibinfo{person}{R.~Yu}} (Eds.), Vol.~\bibinfo{volume}{2025}.
  \bibinfo{pages}{65882--65919}.
\newblock
\urldef\tempurl\url{https://proceedings.iclr.cc/paper_files/paper/2025/file/a4b6ad6b48850c0c331d1259fc66a69c-Paper-Conference.pdf}
\showURL{\tempurl}


\bibitem[\protect\citeauthoryear{Oakes, Yang, Zhou, Houck, Harter,
  Arpaci-Dusseau, and Arpaci-Dusseau}{Oakes et~al\mbox{.}}{2018}]{sock}
\bibfield{author}{\bibinfo{person}{Edward Oakes}, \bibinfo{person}{Leon Yang},
  \bibinfo{person}{Dennis Zhou}, \bibinfo{person}{Kevin Houck},
  \bibinfo{person}{Tyler Harter}, \bibinfo{person}{Andrea Arpaci-Dusseau},
  {and} \bibinfo{person}{Remzi Arpaci-Dusseau}.}
  \bibinfo{year}{2018}\natexlab{}.
\newblock \showarticletitle{{SOCK}: Rapid Task Provisioning with
  {Serverless-Optimized} Containers}. In \bibinfo{booktitle}{\emph{2018
  {USENIX} Annual Technical Conference ({ATC})}}. \bibinfo{publisher}{{USENIX}
  Association}, \bibinfo{pages}{57--70}.
\newblock
\showISBNx{978-1-931971-44-7}
\urldef\tempurl\url{https://www.usenix.org/conference/atc18/presentation/oakes}
\showURL{\tempurl}


\bibitem[\protect\citeauthoryear{{Anthropic}}{{Anthropic}}{[n.\,d.]}]{claudecode-worktrees}
\bibfield{author}{\bibinfo{person}{{Anthropic}}.}
  \bibinfo{year}{[n.\,d.]}\natexlab{}.
\newblock \bibinfo{title}{Best practices for {Claude Code}}.
\newblock \bibinfo{howpublished}{Claude Code Documentation}.
\newblock
\urldef\tempurl\url{https://code.claude.com/docs/en/best-practices}
\showURL{\tempurl}
\newblock
\shownote{Accessed: 2026-10-16}.


\bibitem[\protect\citeauthoryear{Wei, Lu, Wang, Gu, Yang, Chen, and Chen}{Wei
  et~al\mbox{.}}{2023}]{288627}
\bibfield{author}{\bibinfo{person}{Xingda Wei}, \bibinfo{person}{Fangming Lu},
  \bibinfo{person}{Tianxia Wang}, \bibinfo{person}{Jinyu Gu},
  \bibinfo{person}{Yuhan Yang}, \bibinfo{person}{Rong Chen}, {and}
  \bibinfo{person}{Haibo Chen}.} \bibinfo{year}{2023}\natexlab{}.
\newblock \showarticletitle{No Provisioned Concurrency: Fast {RDMA-codesigned}
  Remote Fork for Serverless Computing}. In \bibinfo{booktitle}{\emph{17th
  USENIX Symposium on Operating Systems Design and Implementation (OSDI 23)}}.
  \bibinfo{publisher}{USENIX Association}, \bibinfo{address}{Boston, MA},
  \bibinfo{pages}{497--517}.
\newblock
\showISBNx{978-1-939133-34-2}
\urldef\tempurl\url{https://www.usenix.org/conference/osdi23/presentation/wei-rdma}
\showURL{\tempurl}


\bibitem[\protect\citeauthoryear{{The {CRIU} Project}}{{The {CRIU}
  Project}}{2016}]{criu-incremental-dump}
\bibfield{author}{\bibinfo{person}{{The {CRIU} Project}}.}
  \bibinfo{year}{2016}\natexlab{}.
\newblock \bibinfo{title}{Incremental dumps}.
\newblock \bibinfo{howpublished}{\url{https://criu.org/Incremental_dumps}}.
\newblock


\bibitem[\protect\citeauthoryear{Ding, Wang, Shen, Chen, and Chen}{Ding
  et~al\mbox{.}}{2023}]{288554}
\bibfield{author}{\bibinfo{person}{Haoran Ding}, \bibinfo{person}{Zhaoguo
  Wang}, \bibinfo{person}{Zhuohao Shen}, \bibinfo{person}{Rong Chen}, {and}
  \bibinfo{person}{Haibo Chen}.} \bibinfo{year}{2023}\natexlab{}.
\newblock \showarticletitle{Automated Verification of Idempotence for Stateful
  Serverless Applications}. In \bibinfo{booktitle}{\emph{17th USENIX Symposium
  on Operating Systems Design and Implementation (OSDI 23)}}.
  \bibinfo{publisher}{USENIX Association}, \bibinfo{address}{Boston, MA},
  \bibinfo{pages}{887--910}.
\newblock
\showISBNx{978-1-939133-34-2}
\urldef\tempurl\url{https://www.usenix.org/conference/osdi23/presentation/ding}
\showURL{\tempurl}


\bibitem[\protect\citeauthoryear{Elnozahy, Alvisi, Wang, and Johnson}{Elnozahy
  et~al\mbox{.}}{2002}]{10.1145/568522.568525}
\bibfield{author}{\bibinfo{person}{E.~N.~(Mootaz) Elnozahy},
  \bibinfo{person}{Lorenzo Alvisi}, \bibinfo{person}{Yi-Min Wang}, {and}
  \bibinfo{person}{David~B. Johnson}.} \bibinfo{year}{2002}\natexlab{}.
\newblock \showarticletitle{A survey of rollback-recovery protocols in
  message-passing systems}.
\newblock \bibinfo{journal}{\emph{ACM Comput. Surv.}} \bibinfo{volume}{34},
  \bibinfo{number}{3} (\bibinfo{date}{Sept.} \bibinfo{year}{2002}),
  \bibinfo{pages}{375–408}.
\newblock
\showISSN{0360-0300}
\urldef\tempurl\url{https://doi.org/10.1145/568522.568525}
\showDOI{\tempurl}


\bibitem[\protect\citeauthoryear{{The {CRIU} Project}}{{The {CRIU}
  Project}}{2019}]{criu-userfaultfd}
\bibfield{author}{\bibinfo{person}{{The {CRIU} Project}}.}
  \bibinfo{year}{2019}\natexlab{}.
\newblock \bibinfo{title}{Userfaultfd}.
\newblock \bibinfo{howpublished}{\url{https://criu.org/Userfaultfd}}.
\newblock


\bibitem[\protect\citeauthoryear{{LangChain, Inc.}}{{LangChain, Inc.}}{2024}]{langgraph}
\bibfield{author}{\bibinfo{person}{{LangChain, Inc.}}}
  \bibinfo{year}{2024}\natexlab{}.
\newblock \bibinfo{title}{{LangGraph}: Building Stateful, Multi-Actor
  Applications with {LLMs}}.
\newblock
  \bibinfo{howpublished}{\url{https://github.com/langchain-ai/langgraph}}.
\newblock


\bibitem[\protect\citeauthoryear{Kwon, Li, Zhuang, Sheng, Zheng, Yu, Gonzalez,
  Zhang, and Stoica}{Kwon et~al\mbox{.}}{2023}]{vllm}
\bibfield{author}{\bibinfo{person}{Woosuk Kwon}, \bibinfo{person}{Zhuohan Li},
  \bibinfo{person}{Siyuan Zhuang}, \bibinfo{person}{Ying Sheng},
  \bibinfo{person}{Lianmin Zheng}, \bibinfo{person}{Cody~Hao Yu},
  \bibinfo{person}{Joseph~E. Gonzalez}, \bibinfo{person}{Hao Zhang}, {and}
  \bibinfo{person}{Ion Stoica}.} \bibinfo{year}{2023}\natexlab{}.
\newblock \showarticletitle{Efficient Memory Management for Large Language
  Model Serving with {PagedAttention}}. In
  \bibinfo{booktitle}{\emph{Proceedings of the ACM SIGOPS 29th Symposium on
  Operating Systems Principles}} \emph{(\bibinfo{series}{SOSP '23})}.
\newblock


\bibitem[\protect\citeauthoryear{Sheng, Zhang, Ye, Wu, Zhang, Zhang, Peng, Lin,
  and Wu}{Sheng et~al\mbox{.}}{2025}]{verl}
\bibfield{author}{\bibinfo{person}{Guangming Sheng}, \bibinfo{person}{Chi
  Zhang}, \bibinfo{person}{Zilingfeng Ye}, \bibinfo{person}{Xibin Wu},
  \bibinfo{person}{Wang Zhang}, \bibinfo{person}{Ru Zhang},
  \bibinfo{person}{Yanghua Peng}, \bibinfo{person}{Haibin Lin}, {and}
  \bibinfo{person}{Chuan Wu}.} \bibinfo{year}{2025}\natexlab{}.
\newblock \showarticletitle{{HybridFlow}: A Flexible and Efficient {RLHF}
  Framework}. In \bibinfo{booktitle}{\emph{Proceedings of the Twentieth
  European Conference on Computer Systems}} \emph{(\bibinfo{series}{EuroSys
  '25})}.
\newblock


\bibitem[\protect\citeauthoryear{Hu, Wu, Shen, Liu, Zhu, Wang, Jiang, Wang,
  Chen, Chen, Fang, Xianyu, Cao, Xu, and Liu}{Hu et~al\mbox{.}}{2025}]{openrlhf}
\bibfield{author}{\bibinfo{person}{Jian Hu}, \bibinfo{person}{Xibin Wu},
  \bibinfo{person}{Wei Shen}, \bibinfo{person}{Jason~Klein Liu},
  \bibinfo{person}{Zilin Zhu}, \bibinfo{person}{Weixun Wang},
  \bibinfo{person}{Songlin Jiang}, \bibinfo{person}{Haoran Wang},
  \bibinfo{person}{Hao Chen}, \bibinfo{person}{Bin Chen},
  \bibinfo{person}{Weikai Fang}, \bibinfo{person}{Xianyu}, \bibinfo{person}{Yu
  Cao}, \bibinfo{person}{Haotian Xu}, {and} \bibinfo{person}{Yiming Liu}.}
  \bibinfo{year}{2025}\natexlab{}.
\newblock \bibinfo{title}{{OpenRLHF}: An Easy-to-use, Scalable and
  High-performance {RLHF} Framework}.
\newblock
\newblock
\showeprint[arxiv]{2405.11143}~[cs.AI]
\urldef\tempurl\url{https://arxiv.org/abs/2405.11143}
\showURL{\tempurl}


\bibitem[\protect\citeauthoryear{{ZeroBoot}}{{ZeroBoot}}{2026}]{zeroboot}
\bibfield{author}{\bibinfo{person}{{ZeroBoot}}.}
  \bibinfo{year}{2026}\natexlab{}.
\newblock \bibinfo{title}{{ZeroBoot}: Sub-millisecond {VM} Sandboxes for {AI}
  Agents via Copy-on-Write Forking}.
\newblock
  \bibinfo{howpublished}{\url{https://github.com/zerobootdev/zeroboot}}.
\newblock


\bibitem[\protect\citeauthoryear{Xu, Zhou, Wu, and Kaffes}{Xu
  et~al\mbox{.}}{2025}]{xu2025systemsfoundationsagenticexploration}
\bibfield{author}{\bibinfo{person}{Jiakai Xu}, \bibinfo{person}{Tianle Zhou},
  \bibinfo{person}{Eugene Wu}, {and} \bibinfo{person}{Kostis Kaffes}.}
  \bibinfo{year}{2025}\natexlab{}.
\newblock \bibinfo{title}{Toward Systems Foundations for Agentic Exploration}.
\newblock
\newblock
\showeprint[arxiv]{2510.05556}~[cs.DC]
\urldef\tempurl\url{https://arxiv.org/abs/2510.05556}
\showURL{\tempurl}


\bibitem[\protect\citeauthoryear{Wang and Zheng}{Wang and Zheng}{2026}]{wang2026forkexplorecommitos}
\bibfield{author}{\bibinfo{person}{Cong Wang} {and} \bibinfo{person}{Yusheng
  Zheng}.} \bibinfo{year}{2026}\natexlab{}.
\newblock \bibinfo{title}{Fork, Explore, Commit: OS Primitives for Agentic
  Exploration}.
\newblock
\newblock
\showeprint[arxiv]{2602.08199}~[cs.OS]
\urldef\tempurl\url{https://arxiv.org/abs/2602.08199}
\showURL{\tempurl}


\bibitem[\protect\citeauthoryear{Wu, Chang, Cao, Gao, and Wang}{Wu
  et~al\mbox{.}}{2026}]{wu2026crabsemanticsawarecheckpointrestoreruntime}
\bibfield{author}{\bibinfo{person}{Tianyuan Wu}, \bibinfo{person}{Chaokun
  Chang}, \bibinfo{person}{Lunxi Cao}, \bibinfo{person}{Wei Gao}, {and}
  \bibinfo{person}{Wei Wang}.} \bibinfo{year}{2026}\natexlab{}.
\newblock \bibinfo{title}{Crab: A Semantics-Aware Checkpoint/Restore Runtime
  for Agent Sandboxes}.
\newblock
\newblock
\showeprint[arxiv]{2604.28138}~[cs.OS]
\urldef\tempurl\url{https://arxiv.org/abs/2604.28138}
\showURL{\tempurl}


\bibitem[\protect\citeauthoryear{Huang et~al\mbox{.}}{Huang
  et~al\mbox{.}}{2026}]{huang2026deepseekelasticcomputedsec}
\bibfield{author}{\bibinfo{person}{Jialiang Huang} {et~al\mbox{.}}}
  \bibinfo{year}{2026}\natexlab{}.
\newblock \bibinfo{title}{DeepSeek Elastic Compute (DSec): A Sandbox
  Infrastructure for Effective Agentic Training at Scale}.
\newblock
\newblock
\showeprint[arxiv]{2609.22978}~[cs.DC]
\urldef\tempurl\url{https://arxiv.org/abs/2609.22978}
\showURL{\tempurl}


\bibitem[\protect\citeauthoryear{Cadden, Unger, Awad, Dong, Krieger, and
  Appavoo}{Cadden et~al\mbox{.}}{2020}]{seuss}
\bibfield{author}{\bibinfo{person}{James Cadden}, \bibinfo{person}{Thomas
  Unger}, \bibinfo{person}{Yara Awad}, \bibinfo{person}{Han Dong},
  \bibinfo{person}{Orran Krieger}, {and} \bibinfo{person}{Jonathan Appavoo}.}
  \bibinfo{year}{2020}\natexlab{}.
\newblock \showarticletitle{SEUSS: skip redundant paths to make serverless
  fast}. In \bibinfo{booktitle}{\emph{Proceedings of the Fifteenth European
  Conference on Computer Systems}} (Heraklion, Greece)
  \emph{(\bibinfo{series}{EuroSys '20})}. \bibinfo{publisher}{Association for
  Computing Machinery}, \bibinfo{address}{New York, NY, USA}, Article
  \bibinfo{articleno}{32}, \bibinfo{numpages}{15}~pages.
\newblock
\showISBNx{9781450368827}
\urldef\tempurl\url{https://doi.org/10.1145/3342195.3392698}
\showDOI{\tempurl}


\bibitem[\protect\citeauthoryear{Holmes, Dinis, Honcharuk, Fried, and
  Belay}{Holmes et~al\mbox{.}}{2025}]{spice}
\bibfield{author}{\bibinfo{person}{Ben Holmes}, \bibinfo{person}{Baltasar
  Dinis}, \bibinfo{person}{Lana Honcharuk}, \bibinfo{person}{Joshua Fried},
  {and} \bibinfo{person}{Adam Belay}.} \bibinfo{year}{2025}\natexlab{}.
\newblock \bibinfo{title}{Taming Serverless Cold Starts Through {OS}
  Co-Design}.
\newblock
\newblock
\showeprint[arxiv]{2509.14292}~[cs.OS]
\urldef\tempurl\url{https://arxiv.org/abs/2509.14292}
\showURL{\tempurl}


\bibitem[\protect\citeauthoryear{Kuchler, Li, Zhang, Cvetkovi{\'c}, Goranov,
  Stocker, Thomm, Kalbermatter, Notter, Lattuada, and Klimovic}{Kuchler
  et~al\mbox{.}}{2025}]{10.1145/3731569.3764803}
\bibfield{author}{\bibinfo{person}{Tom Kuchler}, \bibinfo{person}{Pinghe Li},
  \bibinfo{person}{Yazhuo Zhang}, \bibinfo{person}{Lazar Cvetkovi{\'c}},
  \bibinfo{person}{Boris Goranov}, \bibinfo{person}{Tobias Stocker},
  \bibinfo{person}{Leon Thomm}, \bibinfo{person}{Simone Kalbermatter},
  \bibinfo{person}{Tim Notter}, \bibinfo{person}{Andrea Lattuada}, {and}
  \bibinfo{person}{Ana Klimovic}.} \bibinfo{year}{2025}\natexlab{}.
\newblock \showarticletitle{Unlocking True Elasticity for the Cloud-Native Era
  with Dandelion}. In \bibinfo{booktitle}{\emph{Proceedings of the ACM SIGOPS
  31st Symposium on Operating Systems Principles}} (Seoul, Republic of Korea)
  \emph{(\bibinfo{series}{SOSP '25})}. \bibinfo{publisher}{Association for
  Computing Machinery}, \bibinfo{address}{New York, NY, USA},
  \bibinfo{pages}{944–961}.
\newblock
\showISBNx{9798400718700}
\urldef\tempurl\url{https://doi.org/10.1145/3731569.3764803}
\showDOI{\tempurl}


\bibitem[\protect\citeauthoryear{Yang, Du, Song, and Xia}{Yang
  et~al\mbox{.}}{2024}]{10.1145/3698038.3698510}
\bibfield{author}{\bibinfo{person}{Yanning Yang}, \bibinfo{person}{Dong Du},
  \bibinfo{person}{Haitao Song}, {and} \bibinfo{person}{Yubin Xia}.}
  \bibinfo{year}{2024}\natexlab{}.
\newblock \showarticletitle{On-demand and Parallel Checkpoint/Restore for GPU
  Applications}. In \bibinfo{booktitle}{\emph{Proceedings of the 2024 ACM
  Symposium on Cloud Computing}} (Redmond, WA, USA)
  \emph{(\bibinfo{series}{SoCC '24})}. \bibinfo{publisher}{Association for
  Computing Machinery}, \bibinfo{address}{New York, NY, USA},
  \bibinfo{pages}{415–433}.
\newblock
\showISBNx{9798400712869}
\urldef\tempurl\url{https://doi.org/10.1145/3698038.3698510}
\showDOI{\tempurl}


\bibitem[\protect\citeauthoryear{Tullmann, Lepreau, Ford, and Hibler}{Tullmann
  et~al\mbox{.}}{1996}]{557874}
\bibfield{author}{\bibinfo{person}{P. Tullmann}, \bibinfo{person}{J. Lepreau},
  \bibinfo{person}{B. Ford}, {and} \bibinfo{person}{M. Hibler}.}
  \bibinfo{year}{1996}\natexlab{}.
\newblock \showarticletitle{User-level checkpointing through exportable kernel
  state}. In \bibinfo{booktitle}{\emph{Proceedings of the Fifth International
  Workshop on Object-Orientation in Operation Systems}}.
  \bibinfo{pages}{85--88}.
\newblock
\urldef\tempurl\url{https://doi.org/10.1109/IWOOOS.1996.557874}
\showDOI{\tempurl}


\bibitem[\protect\citeauthoryear{Vogt, Miraglia, Portokalidis, Bos, Tanenbaum,
  and Giuffrida}{Vogt et~al\mbox{.}}{2015}]{10.1145/2814576.2814802}
\bibfield{author}{\bibinfo{person}{Dirk Vogt}, \bibinfo{person}{Armando
  Miraglia}, \bibinfo{person}{Georgios Portokalidis}, \bibinfo{person}{Herbert
  Bos}, \bibinfo{person}{Andy Tanenbaum}, {and} \bibinfo{person}{Cristiano
  Giuffrida}.} \bibinfo{year}{2015}\natexlab{}.
\newblock \showarticletitle{Speculative Memory Checkpointing}. In
  \bibinfo{booktitle}{\emph{Proceedings of the 16th Annual Middleware
  Conference}} (Vancouver, BC, Canada) \emph{(\bibinfo{series}{Middleware
  '15})}. \bibinfo{publisher}{Association for Computing Machinery},
  \bibinfo{address}{New York, NY, USA}, \bibinfo{pages}{197–209}.
\newblock
\showISBNx{9781450336185}
\urldef\tempurl\url{https://doi.org/10.1145/2814576.2814802}
\showDOI{\tempurl}


\bibitem[\protect\citeauthoryear{Dearle and Hulse}{Dearle and Hulse}{1995}]{470583}
\bibfield{author}{\bibinfo{person}{A. Dearle} {and} \bibinfo{person}{D.
  Hulse}.} \bibinfo{year}{1995}\natexlab{}.
\newblock \showarticletitle{On page-based optimistic process checkpointing}. In
  \bibinfo{booktitle}{\emph{Proceedings of International Workshop on Object
  Orientation in Operating Systems}}. \bibinfo{pages}{24--32}.
\newblock
\urldef\tempurl\url{https://doi.org/10.1109/IWOOS.1995.470583}
\showDOI{\tempurl}


\bibitem[\protect\citeauthoryear{Tsalapatis, Hancock, Barnes, and
  Mashtizadeh}{Tsalapatis et~al\mbox{.}}{2021}]{10.1145/3477132.3483563}
\bibfield{author}{\bibinfo{person}{Emil Tsalapatis}, \bibinfo{person}{Ryan
  Hancock}, \bibinfo{person}{Tavian Barnes}, {and}
  \bibinfo{person}{Ali~Jos\'{e} Mashtizadeh}.} \bibinfo{year}{2021}\natexlab{}.
\newblock \showarticletitle{The Aurora Single Level Store Operating System}. In
  \bibinfo{booktitle}{\emph{Proceedings of the ACM SIGOPS 28th Symposium on
  Operating Systems Principles}} (Virtual Event, Germany)
  \emph{(\bibinfo{series}{SOSP '21})}. \bibinfo{publisher}{Association for
  Computing Machinery}, \bibinfo{address}{New York, NY, USA},
  \bibinfo{pages}{788–803}.
\newblock
\showISBNx{9781450387095}
\urldef\tempurl\url{https://doi.org/10.1145/3477132.3483563}
\showDOI{\tempurl}


\bibitem[\protect\citeauthoryear{Plank, Beck, Kingsley, and Li}{Plank
  et~al\mbox{.}}{1995}]{10.5555/1267411.1267429}
\bibfield{author}{\bibinfo{person}{James~S. Plank}, \bibinfo{person}{Micah
  Beck}, \bibinfo{person}{Gerry Kingsley}, {and} \bibinfo{person}{Kai Li}.}
  \bibinfo{year}{1995}\natexlab{}.
\newblock \showarticletitle{Libckpt: transparent checkpointing under Unix}. In
  \bibinfo{booktitle}{\emph{Proceedings of the USENIX 1995 Technical Conference
  Proceedings}} (New Orleans, Louisiana) \emph{(\bibinfo{series}{TCON'95})}.
  \bibinfo{publisher}{USENIX Association}, \bibinfo{address}{USA},
  \bibinfo{pages}{18}.
\newblock


\bibitem[\protect\citeauthoryear{Ansel, Arya, and Cooperman}{Ansel
  et~al\mbox{.}}{2009}]{dmtcp}
\bibfield{author}{\bibinfo{person}{Jason Ansel}, \bibinfo{person}{Kapil Arya},
  {and} \bibinfo{person}{Gene Cooperman}.} \bibinfo{year}{2009}\natexlab{}.
\newblock \showarticletitle{DMTCP: Transparent checkpointing for cluster
  computations and the desktop}. In \bibinfo{booktitle}{\emph{Proceedings of
  the 2009 IEEE International Symposium on Parallel and Distributed
  Processing}} \emph{(\bibinfo{series}{IPDPS '09})}. \bibinfo{publisher}{IEEE
  Computer Society}, \bibinfo{address}{USA}, \bibinfo{pages}{1--12}.
\newblock
\showISBNx{9781424437511}
\urldef\tempurl\url{https://doi.org/10.1109/IPDPS.2009.5161063}
\showDOI{\tempurl}


\bibitem[\protect\citeauthoryear{Lazarev, Gohil, Tsai, Anderson, Chitlur,
  Zhang, and Delimitrou}{Lazarev et~al\mbox{.}}{2024}]{sabre}
\bibfield{author}{\bibinfo{person}{Nikita Lazarev}, \bibinfo{person}{Varun
  Gohil}, \bibinfo{person}{James Tsai}, \bibinfo{person}{Andy Anderson},
  \bibinfo{person}{Bhushan Chitlur}, \bibinfo{person}{Zhiru Zhang}, {and}
  \bibinfo{person}{Christina Delimitrou}.} \bibinfo{year}{2024}\natexlab{}.
\newblock \showarticletitle{Sabre: hardware-accelerated snapshot compression
  for serverless MicroVMs}. In \bibinfo{booktitle}{\emph{Proceedings of the
  18th USENIX Conference on Operating Systems Design and Implementation}}
  (Santa Clara, CA, USA) \emph{(\bibinfo{series}{OSDI'24})}.
  \bibinfo{publisher}{USENIX Association}, \bibinfo{address}{USA}, Article
  \bibinfo{articleno}{1}, \bibinfo{numpages}{18}~pages.
\newblock
\showISBNx{978-1-939133-40-3}


\bibitem[\protect\citeauthoryear{{The Btrfs Project}}{{The Btrfs
  Project}}{2009}]{btrfs}
\bibfield{author}{\bibinfo{person}{{The Btrfs Project}}.}
  \bibinfo{year}{2009}\natexlab{}.
\newblock \bibinfo{title}{{Btrfs} Documentation}.
\newblock \bibinfo{howpublished}{\url{https://btrfs.readthedocs.io}}.
\newblock


\bibitem[\protect\citeauthoryear{{OpenZFS}}{{OpenZFS}}{2013}]{zfs}
\bibfield{author}{\bibinfo{person}{{OpenZFS}}.}
  \bibinfo{year}{2013}\natexlab{}.
\newblock \bibinfo{title}{{OpenZFS} Documentation}.
\newblock
  \bibinfo{howpublished}{\url{https://openzfs.github.io/openzfs-docs/}}.
\newblock


\bibitem[\protect\citeauthoryear{{Linux Kernel Project}}{{Linux Kernel
  Project}}{2018}]{erofs}
\bibfield{author}{\bibinfo{person}{{Linux Kernel Project}}.}
  \bibinfo{year}{2018}\natexlab{}.
\newblock \bibinfo{title}{{EROFS}: Enhanced Read-Only File System}.
\newblock \bibinfo{howpublished}{\url{https://erofs.docs.kernel.org}}.
\newblock


\bibitem[\protect\citeauthoryear{{LangChain, Inc.}}{{LangChain, Inc.}}{2022}]{langchain}
\bibfield{author}{\bibinfo{person}{{LangChain, Inc.}}}
  \bibinfo{year}{2022}\natexlab{}.
\newblock \bibinfo{title}{{LangChain}: Building Applications with {LLMs}
  through Composability}.
\newblock
  \bibinfo{howpublished}{\url{https://github.com/langchain-ai/langchain}}.
\newblock


\end{thebibliography}
\end{document}